\documentclass[12pt]{article}
\pdfoutput=1
\usepackage[utf8]{inputenc}
\usepackage{imakeidx}

\usepackage[sectionbib]{natbib}
\usepackage{array,epsfig,rotating}
\usepackage{color}
\usepackage[dvipsnames]{xcolor}
\usepackage{hyperref}
\hypersetup{
  colorlinks,
  linkcolor={red!50!black},
  citecolor={blue},
  urlcolor={blue!80!black}
}

\usepackage[ruled,vlined]{algorithm2e}
\usepackage{tabularx}
\usepackage{multirow}
\usepackage{booktabs}
\usepackage{adjustbox}
\usepackage{amsmath}
\usepackage{amssymb}
\usepackage{amsfonts}
\usepackage{amsthm}
\usepackage{float}
\usepackage{verbatim}
\usepackage{setspace}
\usepackage{caption}
\usepackage{multirow}
\usepackage{blkarray}

\usepackage{graphbox}
\usepackage{bbm}
\allowdisplaybreaks

\usepackage{subcaption}
\usepackage{cleveref} 
\newtheorem{theorem}{Theorem}

\newtheorem{proposition}{Proposition}
\newtheorem{definition}{Definition}
\newtheorem{example}{Example}
\newtheorem{remark}{Remark}

\newcommand{\aaa}{\boldsymbol \alpha}
\newcommand{\bo}{\boldsymbol }

\newcommand{\one}{\mathbf 1}

\newcommand{\pa}{\text{\normalfont{pa}}}

\newcommand{\ma}{\mathbf A}
\newcommand{\QQ}{{\mathbf Q}}

\newcommand{\bg}{{\boldsymbol g}}
\newcommand{\bs}{{\boldsymbol s}}
\newcommand{\bq}{{\boldsymbol q}}
\newcommand{\bt}{{\boldsymbol t}}
\newcommand{\pp}{{\boldsymbol p}}

\newcommand{\RR}{\mathbf R}
\newcommand{\GG}{\mathbf G}
\newcommand{\EE}{\mathbf E}
\newcommand{\PP}{\mathbf P}

\newcommand{\TT}{\boldsymbol \Theta}

\newcommand{\mca}{\mathcal A}
\newcommand{\mce}{\mathcal E}
\def\hat{\widehat}

\usepackage[margin=1in]{geometry}

\usepackage{tikz}
\usetikzlibrary{shapes, calc, shapes, arrows, positioning}

\tikzstyle{qedge}=[->,thick,black]
\tikzstyle{pre}=[->,thick,dotted]
\tikzstyle{pres}=[-,dotted]

\definecolor{myblue}{rgb}{0.0265,    0.6137,    0.8135}
\definecolor{myyellow}{rgb}{0.9290,    0.6940,    0.1250}

\newcommand{\darkblue}[1]{#1}

\usepackage{bbm}
\tikzstyle{neuron}=[draw, circle,minimum size=25pt,inner sep=0pt, fill=black!20]

\tikzstyle{hidden}=[draw, circle,minimum size=25pt,inner sep=0pt, fill=white]

\tikzstyle{hiddens}=[draw,circle,minimum size=17pt,inner sep=0pt, fill=white]

\usetikzlibrary{arrows.meta}
\tikzset{>={Latex[width=2mm,length=2mm]}}

\tikzstyle{arr}=[->, thick, black]

\usetikzlibrary{decorations.text}

\usetikzlibrary{shadows,shadings,shapes.symbols}
\tikzset{
    double color fill/.code 2 args={
        \pgfdeclareverticalshading[%
            tikz@axis@top,tikz@axis@middle,tikz@axis@bottom%
        ]{diagonalfill}{100bp}{%
            color(0bp)=(tikz@axis@bottom);
            color(50bp)=(tikz@axis@bottom);
            color(50bp)=(tikz@axis@middle);
            color(50bp)=(tikz@axis@top);
            color(100bp)=(tikz@axis@top)
        }
        \tikzset{shade, left color=#1, right color=#2, shading=diagonalfill}
    }
}

\def\spacingset#1{\renewcommand{\baselinestretch}%
{#1}\small\normalsize} 
\spacingset{1.45}

\newcommand\blfootnote[1]{%
  \begingroup
  \renewcommand\thefootnote{}\footnote{#1}%
  \addtocounter{footnote}{-1}%
  \endgroup
}

\title{Latent Conjunctive Bayesian Network: Unify Attribute Hierarchy and Bayesian Network for Cognitive Diagnosis}
\author{\large Seunghyun Lee$^\dagger$ and Yuqi Gu$^*$}
\date{\large Department of Statistics, Columbia University}

\begin{document}
\spacingset{1}
\maketitle
\blfootnote{$^\dagger$\texttt{sl4963@columbia.edu.}}
\blfootnote{$^*$\texttt{yuqi.gu@columbia.edu.} This work is partially supported by NSF Grant DMS-2210796.}

\vspace{-5mm}
\begin{abstract}
Cognitive diagnostic assessment aims to measure specific knowledge
structures in students. To model data arising from such assessments, cognitive diagnostic models with discrete latent variables have gained popularity in educational and behavioral sciences. In a learning context, the latent variables often denote sequentially acquired skill attributes, which is often modeled by the so-called attribute hierarchy method.
One drawback of the traditional  attribute hierarchy method is that its parameter complexity varies substantially with the hierarchy’s graph structure, lacking statistical parsimony. Additionally, arrows among the attributes do not carry an interpretation of statistical dependence. Motivated by these, we propose a new family of \emph{latent conjunctive Bayesian networks} (LCBNs), which rigorously unify the attribute hierarchy method for sequential skill mastery and the Bayesian network model  in statistical machine learning. In an LCBN, the latent graph not only retains the hard constraints on skill prerequisites as an attribute hierarchy, but also encodes nice conditional independence interpretation as a Bayesian network. LCBNs are identifiable, interpretable, and parsimonious statistical tools to diagnose students’ cognitive abilities from assessment data. We propose an efficient two-step EM algorithm for structure learning and parameter estimation in LCBNs, and establish the consistency of this procedure.
Application of our method to an international educational assessment dataset gives interpretable findings of cognitive diagnosis.
\end{abstract}


\spacingset{1.45}
\section{Introduction}
Cognitive diagnostic assessment aims to measure specific knowledge structures and processing skills in students \citep{leighton2007cdabook}.
To model data arising from such assessments, \emph{cognitive diagnostic models} (CDMs) with discrete latent variables \citep[also called diagnostic classification models; see][]{rupp2010diagnostic, von2019handbook} have recently gained great popularity in educational, psychological, and behavioral applications. 

CDMs adopt a set of discrete latent \emph{attributes} with substantive meaning to explain a subject's multivariate responses to a set of items.
``Attribute'' here is a generic term that can represent unobserved psychological constructs including skills, knowledge states, conceptual understandings, cognitive processes, and rules \citep{wang2021tcdm}.
In educational settings, each attribute often represents the mastery/deficiency of a specific latent skill.
Adopting CDMs in educational assessment can generate fine-grained diagnoses about students' multiple latent skills, and hence provide detailed feedback about their weaknesses and strengths.
A typical CDM consists of a \emph{structural model} for the latent attributes and a \emph{measurement model} describing the dependence of the observed variables (i.e., item responses in educational assessments) on the latent attributes.
The measurement model is accompanied by a so-called $\QQ$-matrix \citep{tatsuoka1983rule}, summarizing which subset of the attributes each observed variable measures or requires. The $\QQ$-matrix is often pre-specified by domain experts.

Various measurement models have been proposed for different diagnostic purposes.
For example, the popular and fundamental Deterministic Input Noisy Output ``AND'' gate \citep[DINA;][]{junker2001cognitive} model adopts the conjunctive assumption by specifying that a student needs to master all attributes required by an item to be capable of it.
The generalized DINA \citep[GDINA;][]{de2011generalized} model generalizes this by incorporating main effects and interaction effects of required attributes into the measurement model. 
Other popular CDMs include the Deterministic Input Noisy Output ``OR'' gate \citep[DINO;][]{templin2006measurement} model, the log-linear CDM \citep[LCDM;][]{henson2009lcdm}, the additive CDM \citep[ACDM;][]{de2011generalized}, and general diagnostic models \citep[GDM;][]{von2008general}.

As for the structural model for the latent attributes in a CDM, the \textit{attribute hierarchy method} that models sequential skill mastery has recently attracted increasing attention \citep{leighton2004attribute, gierl2007ahm, wang2011ahm, templin2014hierarchical, gu2019jmlr, wang2021jebs}.
Students' learning is not instantaneous and often proceeds in a sequential and dependent manner.
In a learning context, possessing lower level skills are often believed to be the
prerequisite for possessing higher level skills \citep{simon2012explicating, briggs2012psychometric}.
\cite{leighton2004attribute} first proposed the attribute hierarchy method, and \cite{templin2014hierarchical}  integrated the attribute hierarchy with a flexible measurement model in a statistical framework to define the family of hierarchical cognitive diagnostic models (HCDMs). HCDMs adopt the \emph{unstructured statistical model} for the attribute patterns under a hierarchy. Specifically, in an HCDM, each pattern respecting the attribute hierarchy has an unstructured proportion parameter, which characterizes how much proportion of the student population possess this skill pattern.

Most existing studies on attribute hierarchy followed  \cite{templin2014hierarchical} to adopt the unstructured model for hierarchies. One limitation of this popular approach is that its parameter complexity varies substantially with the graph structure of the hierarchy, lacking statistical parsimony. For instance, with $K$ binary attributes, a chain graph hierarchy requires $K$ free parameters for the latent distribution, whereas a graph with one attribute serving as a common parent to all the other attributes requires $2^{K-1}$ parameters. This lack of parsimony especially creates computational and statistical challenges when there are a large number of attributes and a limited sample size.
In addition, the unstructured model for attribute hierarchy does not endow the hierarchy graph with any probabilistic interpretation.
Specifically, the hierarchy among the latent attributes is merely treated as a machinery for inducing hard constraints on which latent attribute patterns are permissible (those respecting the hierarchy) and which are forbidden (those violating the hierarchy). As a result, the arrows in such an attribute hierarchy graph do not carry clear interpretation of direct statistical dependence, nor does the lack of arrows indicate conditional independence. 

Motivated by the above issues, we propose a new family of \emph{latent conjunctive Bayesian networks} (LCBNs) for cognitive diagnosis. LCBNs are a parsimonious and interpretable class of probabilistic graphical models that rigorously unify attribute hierarchy and Bayesian network.
A Bayesian network \citep{pearl1988probabilistic} is a directed graphical model of random variables, in which directed arrows indicate statistical dependence and the lack of arrows indicate conditional independence.
In our LCBN, the directed acyclic graph among the latent attributes \emph{not only} respects the hard constraints on which attribute patterns are permissible/forbidden as under a usual attribute hierarchy, \emph{but also} encodes the nice conditional independence interpretation as in a usual Bayesian network.
Therefore, LCBNs enjoy the best of both worlds. Moveover, LCBNs are parsimonious statistical models with a fixed parameter complexity in the latent part -- it always only requires $K$ parameters for specifying the joint distribution of $K$ binary latent attributes, regardless of the graph structure of the hierarchy. 

In terms of model identifiability, we prove that the attribute hierarchy graph and all the continuous parameters in an LCBN are fully identifiable from the observed data distribution. Our identifiability conditions are transparent requirements on the discrete structure in the model.
Identifiability lays the foundation for valid statistical estimation and inference. In terms of estimation, we propose an efficient two-step EM algorithm to perform structure learning and parameter estimation in LCBNs. In the first step, we leverage a penalized EM algorithm for selecting significant latent patterns \citep{gu2019jmlr} to estimate the discrete structure -- the attribute hierarchy graph. In the second step, we fix the attribute hierarchy and propose another EM algorithm to estimate the continuous parameters in the LCBN. Simulation studies demonstrate the estimation accuracy of this procedure. We apply our method to analyze a dataset extracted from an international educational assessment, the  Trends in Mathematics and Science Study (TIMSS). The real data analysis gives interpretable finds of cognitive diagnosis and demonstrates the wide applicability of our method. 

The remainder of this paper is organized as follows.  Section \ref{sec:motivation} introduces the background of cognitive diagostic modeling, proposes the general framework of LCBNs, and discusses some related work. 
Section \ref{sec:id} provides  identifiability conditions of LCBNs. Section  \ref{sec:est} proposes a two-step EM algorithm to estimate the attribute hierarchy graph and model parameters in LCBNs. 
Section \ref{sec:sim} presents simulation studies to empirically assess the proposed method. Section \ref{sec:data} applies the new method to analyze an international educational assessment dataset. Finally, Section \ref{sec:conclusion} provides concluding remarks and discusses future directions.
{We also provide the technical proofs of the theorems, additional identifiability results, and additional simulation studies in the Supplementary Material.}

\section{Latent Conjunctive Bayesian Network}\label{sec:motivation}

\subsection{Cognitive Diagnostic Modeling with an attribute hierarchy}

\begin{figure}[h!]
\centering

\resizebox{0.5\textwidth}{!}{
    \begin{tikzpicture}[scale=2]

    \node (v1)[neuron] at (0, 0) {$R_1$};
    \node (v2)[neuron] at (0.8, 0) {$R_2$};
    \node (v3)[neuron] at (1.6, 0) {$R_3$};
    \node (v4)[neuron] at (2.4, 0) {$\cdots$};
    \node (v5)[neuron] at (3.2, 0) {$\cdots$};
    \node (v6)[neuron] at (4, 0)   {$R_J$};
       
    \node (h1)[hidden] at (0.5, 1.2) {$\alpha_1$};
    \node (h2)[hidden] at (1.5, 1.2) {$\alpha_2$};
    \node (h3)[hidden] at (2.5, 1.2) {$\cdots$};
    \node (h4)[hidden] at (3.5, 1.2) {$\alpha_{K}$};

    \node[anchor=west] (g1) at (3.8, 0.7) {{$\QQ_{J \times K}$}}; 
    \node[anchor=west] (g1) at (0.7, 1.6) {{$\mce=\{\alpha_1\to\alpha_2,~ \ldots, ~ \alpha_{K-1}\to\alpha_K \}$}};

    \draw[qedge] (h1) -- (v1) node [midway,above=-0.12cm,sloped] {}; 
    \draw[qedge] (h1) -- (v4) node [midway,above=-0.12cm,sloped] {};  
    \draw[qedge] (h2) -- (v2) node [midway,above=-0.12cm,sloped] {}; 
    \draw[qedge] (h2) -- (v4) node [midway,above=-0.12cm,sloped] {}; 
    \draw[qedge] (h3) -- (v3) node [midway,above=-0.12cm,sloped] {}; 
    \draw[qedge] (h3) -- (v5) node [midway,above=-0.12cm,sloped] {}; 
    \draw[qedge] (h2) -- (v6) node [midway,above=-0.12cm,sloped] {}; 
    \draw[qedge] (h3) -- (v6) node [midway,above=-0.12cm,sloped] {}; 
    \draw[qedge] (h4) -- (v6) node [midway,above=-0.12cm,sloped] {}; 
    
    \draw[black, pre] (h1) -- (h2) node [midway,above=-0.12cm,sloped] {}; 
    \draw[black, pre] (h2) -- (h3) node [midway,above=-0.12cm,sloped] {};
    \draw[black, pre] (h3) -- (h4) node [midway,above=-0.12cm,sloped] {}; 

\end{tikzpicture}
}
\caption{Graphical model representation of a cognitive diagnostic model with a linear attribute hierarchy. White nodes are latent attributes, and grey nodes are observed responses. Dotted arrows denote the prerequisite relationship among the latent attributes, and solid arrows denote the conditional dependence structure of the observed responses given the latent attributes.
}
\label{fig-1layer}
\vspace{-4mm}
\end{figure}
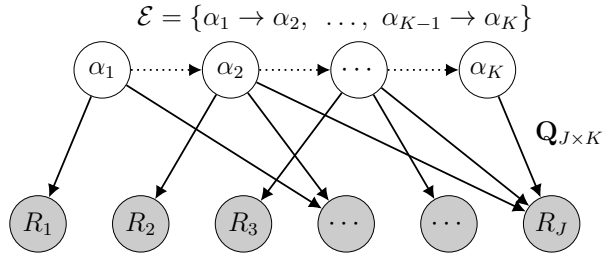

We first introduce the basic setup of a CDM.
Consider a CDM for modeling a cognitive diagnostic assessment. A student's observed variables are his or her correct/wrong  responses  to a set of $J$ items in the  assessment, denoted by $\RR=(R_1,\ldots,R_J) \in\{0,1\}^J$, in which $R_j=1$ indicates the student's response to the $j$th item is correct and $R_j=0$ otherwise. A student's latent variables are his or her profile of  presence/absence of a set of $K$ skill attributes, denoted by $\aaa=(\alpha_1,\ldots,\alpha_K) \in\{0,1\}^K$, in which $\alpha_k=1$ indicates the student masters the $k$th skill and $\alpha_k=0$ otherwise.
Typically, a CDM consists of two parts: a \textit{structural model} for the latent attributes, and a \textit{measurement model} to describe the distribution of the observed responses given the latent. 
In a learning context, the skill attributes are often sequentially acquired and form a hierarchy with prerequisite relations among attributes.
In a CDM with attribute hierarchy, the key elements of the structural and measurement modeling parts are captured by two discrete graph structures: a directed acyclic graph among the latent attributes, and a bipartite directed graph pointing from the latent attributes to the observed responses.  These two graphical structures are illustrated in Figure \ref{fig-1layer}.
For clarity of presentation, we next describe the measurement part and the structural part of a CDM separately in subsequent paragraphs.  

For the measurement part of a CDM, 
educational experts who designed the assessment usually provide information about  which subset of the $K$ skills each test item measures.
All such information are summarized in a so-called $\QQ$-matrix \citep{tatsuoka1983rule}. The $\QQ$-matrix $\QQ = (q_{j,k}) \in \{0, 1\}^{J \times K}$ is a $J\times K$ matrix with binary entries, with rows indexed by observed items and columns by latent attributes. Each entry $q_{j,k} = 1$ or $0$ indicates whether or not the $j$th test item requires/measures the $k$th latent skill.
Consequently, the $j$th row vector of $\QQ$, denoted by $\bq_j=(q_{j,1}, \ldots, q_{j,K})$, is the attribute requirement profile of item $j$.
For example, in Figure \ref{fig-1layer} we have $\bo q_1 = (1, 0, 0, 0)$ since the first item only requires the first attribute.

Statistically, a student's responses to the $J$ items are assumed to be conditionally independent given his or her latent attribute profile $\aaa$. Such a local independence assumption is widely adopted in various models for item response data. 
We collect all the conditional correct response probabilities in a $J\times 2^K$ item parameter matrix $\TT = (\theta_{j, \aaa})_{J\times 2^K}$, with rows indexed by the $J$ test items and columns by the $2^K$ binary pattern configurations in $\{0,1\}^K$. For any $j\in[J]$ and $\aaa\in\{0,1\}^K$, the entry 
$$\theta_{j, \aaa} = \mathbb{P}(R_j = 1 \mid \aaa)
$$ 
defines the conditional probability of giving a correct response to item $j$ given that one has a latent skill profile $\aaa$.
For two vectors $\bo a=(a_1,\ldots,a_K)$ and $\bo b=(b_1,\ldots,b_K)$ of the same length, we write $\bo a\succeq \bo b$ if $a_k\geq b_k$ for all $k\in[K]$ and write $\bo a\nsucceq \bo b$ otherwise.
An important observation is that, since $\bo q_j$ describes which subset of attributes item $j$ measures, the correct response probability $\theta_{j,\aaa}$ only depends on those attributes $\alpha_k$ that are measured by item $j$ (that is, those $\alpha_k$ with $q_{j,k}=1$). Therefore, 
\begin{align}\label{eq-thetaeq}
\theta_{j, \aaa} = \theta_{j, \aaa'} \text{ for any } \darkblue{\aaa, \aaa' \succeq \bq_j}.
\end{align}
Another common feature shared by many different CDM measurement models is that item parameters  often exhibit monotonicity \citep{xu2018identifying, gu2019jmlr, balamuta2022exploratory}:
\begin{equation}\label{eq-mono}
    \theta_{j, \aaa} > \theta_{j, \aaa'} \text{ for any } \aaa \succeq \bq_j \text{ and } \aaa' \nsucceq \bq_j.
\end{equation}
The above inequality can be interpreted as: if a student possesses all the attributes required by item $j$ (that is, $\aaa \succeq \bq_j$), then this student has a higher probability to give a correct response to this item compared to other subjects who lack some required attribute.

We next review some popular and widely used CDM measurement models. 

\begin{example}[DINA model]\label{exp-dina}
The Deterministic Input Noisy output ``And'' gate \citep[DINA;][]{junker2001cognitive} model is a very popular and fundamental CDM. For each item $j$, DINA uses exactly two distinct parameters to describe the conditional distribution of $R_j$. 
Specifically, if a student with latent  profile $\aaa$ masters all the required attributes of item $j$ (i.e., $\aaa\succeq \bo q_j$), then he/she is considered capable of this item but still has a small probability $s_j$ to make a careless mistake; on the other hand, if the student lacks some of the required attributes with $\aaa\nsucceq\bo q_j$, then he/she is considered incapable of this item but still has a small probability $g_j$ to have a lucky guess.
The correct response probability can be written as
\begin{equation}\label{eq:dina}
    \theta_{j, \aaa} = \mathbb{P}(R_j = 1 \mid \aaa) = \begin{cases}
    1-s_j, &\text{ if } \aaa \succeq \bo q_j;\\
    g_j, & \text{ if } \aaa \nsucceq \bo q_j.
    \end{cases}
\end{equation}
$s_j$ and $g_j$ are called slipping parameter and guessing parameter, respectively.
The monotonicity inequality in \eqref{eq-mono} now boils down to $1 - s_j > g_j$ for all $j$. The interpretation is that for any item, a capable student always has a higher probability of giving a correct response than an incapable student.
DINA is widely used in educational cognitive diagnosis due to its parsimony and interpretability.
\end{example}

\begin{example}[Main-effect CDMs]
Main-effect CDMs incorporate the main effects of the latent attributes to model the responses.
Specifically, a main-effect CDM assumes that the probability of $R_j=1$ is a function of the main effects of the attributes required for item $j$. 
\begin{equation}\label{eq-maineff}
    \theta_{j, \aaa} = f(\delta_{j, 0} + \sum_{k = 1}^K \delta_{j, k} q_{j, k} \alpha_k ),
\end{equation}
where $f(\cdot)$ is a monotonic link function.
Note that not all the $\delta_{j,k}$ in the above expression are needed in the model specification. Only when $q_{j,k}=1$ will the corresponding $\delta_{j,k}$ be incorporated in the model. Assuming  $\delta_{j,k}>0$ satisfies the monotonicity requirement \eqref{eq-mono}.
When the link function $f$ is the identity, \eqref{eq-maineff} gives the Additive Cognitive Diagnosis Model \citep[ACDM;][]{de2011generalized}; when $f$ is the inverse logit function, \eqref{eq-maineff} gives the Logistic Linear Model \citep[LLM;][]{maris1999estimating}; yet another parametrization of \eqref{eq-maineff} gives rise to the Reduced Reparameterized Unified Model \citep[R-RUM;][]{dibello1995unified}.

\end{example}

\begin{example}[All-effect CDMs]\label{exp-gdina}
All-effect CDMs generalize both the DINA model and the main-effect CDMs by considering both the main effects and all the interaction effects of the required attributes. The item parameter $\theta_{j, \aaa}$ can be written as
\begin{equation}\label{eq-gdina}
\begin{aligned}
    \theta_{j, \aaa} &= f(\delta_{j, 0} + \sum_{k = 1}^K \delta_{j, k}  q_{j, k} \alpha_k + \sum_{1 \le k < k' \le K} \delta_{j, k k'}  (q_{j,k} \alpha_k) ( q_{j,k'}\alpha_{k'} ) + \cdots + \delta_{j, 1 \cdots K} \prod_{k = 1}^K (q_{j, k} \alpha_k) ),
\end{aligned}
\end{equation}
where $\delta_{j,k}$ is the main effect of the required attribute $\alpha_k$, and $\delta_{j,kk'}$ is the interaction effect between two required attributes $\alpha_k$ and $\alpha_{k'}$, etc.
When the link function $f$ is the identity, \eqref{eq-gdina} gives the Generalized DINA model \citep[GDINA;][]{de2011generalized}; when $f$ is the inverse logit, \eqref{eq-gdina} gives the Log-linear CDM \citep[LCDM;][]{henson2009lcdm}; also see the General Diagnostic Model (GDM) framework in \cite{von2008general}.

\end{example}

We now describe the structural modeling part of a CDM with an attribute hierarchy.
The hierarchy is a collection of prerequisite relations between the $K$ latent attributes, in which possessing lower level, more basic skills are assumed to be the prerequisite for possessing higher level, more advanced ones. For any $1 \le k \neq \ell \le K$, we say that attribute $k$ is a \textit{prerequisite} for attribute $\ell$ (and denote this by $\alpha_k \to \alpha_\ell$ or simply $k\to\ell$) if any latent skill pattern $\aaa\in\{0,1\}^K$ with $\alpha_k = 0$ and $\alpha_\ell = 1$ does not exist in the student population and is not ``permissible''. In other words, for any student that masters the higher level advanced skill $\alpha_\ell$, he/she must have mastered the lower level basic skill $\alpha_k$.
Denote the collection of all the prerequisite relationships by
$$
\mce = \{k\to\ell: \text{ the $k$th skill attribute is a prerequisite for the $\ell$th skill attribute}\}.
$$
Any attribute hierarchy $\mce$ can be visualized as a directed acyclic graph among $K$ nodes, each node representing a latent attribute. For example, Figure \ref{fig-1layer} illustrates a linear hierarchy among the skills with $\mathcal E = \{\alpha_1\to\alpha_2,~ \alpha_2\to\alpha_3,~ \ldots,~ \alpha_{K-1} \to \alpha_K\}$.

Statistically, for a traditional CDM without any attribute hierarchy, the most commonly adopted model for the latent attributes is the unstructured model. This model endows every latent skill profile $\aaa\in\{0,1\}^K$ with a population proportion parameter $p_{\aaa}$, satisfying $p_{\aaa}\geq 0$ and $\sum_{\aaa\in\{0,1\}^K} p_{\aaa} = 1$. The parameter $p_{\aaa}$ describes the proportion in the student population that possesses the attribute pattern $\aaa$.
For a CDM with an attribute hierarchy,
most existing studies followed \cite{templin2014hierarchical} to adopt an unstructured statistical model for the hierarchy.
Specifically, such a model is based on the observation that any nonempty $\mce$ induces a sparsity structure on the $2^K$--dimensional proportion parameters $\pp = (p_{\aaa}: \aaa\in\{0,1\}^K)$. For example, if $k\to\ell$, then as aforementioned, any pattern $\aaa$ with $\alpha_k=0$ but $\alpha_\ell=1$ does not exist in the population and hence its population proportion $p_{\aaa}=0$. In this way, we can define the set of permissible latent skill patterns under a hierarchy $\mathcal E$ as follows:
\begin{equation}\label{eq:ahm def}
    \mca(\mce) 
=
\{\aaa\in\{0,1\}^K:~ \aaa \text{ is permissible under }\mce\}
= 
\{\aaa\in\{0,1\}^K:~ p_{\aaa} > 0 \text{ under } \mce \}.
\end{equation}
Note that $\mca(\mce)$ 
is fully determined by the attribute hierarchy $\mce$. 

Since the hierarchy $\mce$ is a directed acyclic graph among $K$ attributes, it can also be equivalently represented by a $K \times K$ \emph{reachability matrix} $\GG(\mce)$ (also denoted by $\GG$ for short) in the graph theory terminology. The $(k,\ell)$th entry of $\GG$ is a binary indicator of whether the $k$th skill is the prerequisite for the $\ell$th skill, that is, $G_{k,\ell} = \mathbbm{1}(k\to\ell)$. Here we assume the diagonal entries of $\GG$ are all zero. This definition is slightly different from the reachability matrix $\EE$ in \cite{gu2022hlam}, which assumes all diagonal entries to be one. Assuming $\GG$ in our current way is for notational convenience, as to be demonstrated soon in \eqref{eq:LCBN} in the next subsection. The following example illustrates the concepts related to the attribute hierarchy.

\begin{example}\label{ex:toy}
Consider an example with $K = 4$ skill attributes and a hierarchy $\mce = \{ 1 \rightarrow 3,~ 1 \rightarrow 4,~ 2 \rightarrow 3,~ 2 \rightarrow 4\}$. This hierarchy means that the first two skills are the basic ones that serve as the prerequisites for the last two advanced skills. This $\mce$ is visualized in the left panel of Figure \ref{fig-lcbntoy}. There are seven permissible attribute patterns under $\mce$:
\begin{equation}\label{eq-ae}
\mca(\mce) = \{0000,~ 1000,~ 0100,~ 1100,~ 1110,~ 1101,~ 1111 \}.
\end{equation}
The patterns in $\mca(\mce)$ can also be viewed as forming a distributive lattice, a concept in combinatorics \citep{gratzer2009lattice}, as shown in the middle panel of Figure \ref{fig-lcbntoy}. The corresponding reachability matrix $\GG$ under $\mce$ is shown in the rightmost panel of Figure \ref{fig-lcbntoy}.
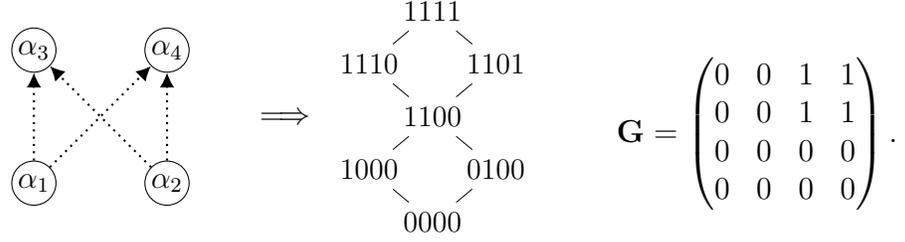
\begin{figure}[h!]
\centering
\qquad\qquad
\begin{minipage}[c]{0.2\textwidth}
\resizebox{0.8\textwidth}{!}{
\begin{tikzpicture}[scale=1.8]
	\node (h1)[hiddens] at (0, 0) {$\alpha_1$};
    \node (h2)[hiddens] at (1, 0) {$\alpha_2$};
    \node (h3)[hiddens] at (0, 1) {$\alpha_3$};
    \node (h4)[hiddens] at (1, 1) {$\alpha_4$};
    
    
    \draw[pre] (h1) -- (h3); \draw[pre] (h1) -- (h4);
    \draw[pre] (h2) -- (h3); \draw[pre] (h2) -- (h4);
\end{tikzpicture}
} 
\end{minipage}
$\Longrightarrow$
\begin{minipage}[c]{0.18\textwidth}
\resizebox{\textwidth}{!}{
\begin{tikzpicture}[scale=1.5]
	\node (p1)[] at (0, 0) {$0000$};
	\node (p2)[] at (-0.6, 0.5) {$1000$};
	\node (p3)[] at (0.6, 0.5) {$0100$};
	\node (p4)[] at (0, 1) {$1100$};
	\node (p5)[] at (-0.6, 1.5) {$1110$};
	\node (p6)[] at (0.6, 1.5) {$1101$};
	\node (p7)[] at (0, 2) {$1111$};
	
	\draw[-] (p1) -- (p2);
	\draw[-] (p1) -- (p3);
	\draw[-] (p2) -- (p4);
	\draw[-] (p3) -- (p4);
	\draw[-] (p4) -- (p5);
	\draw[-] (p4) -- (p6);
	\draw[-] (p5) -- (p7);
	\draw[-] (p6) -- (p7);
	
    
\end{tikzpicture}
} 
\end{minipage}
\qquad
\begin{minipage}[c]{0.2\textwidth}
$$
\GG = \begin{pmatrix}
0 & 0 & 1 & 1 \\
0 & 0 & 1 & 1 \\
0 & 0 & 0 & 0 \\
0 & 0 & 0 & 0 \\
\end{pmatrix}.
$$
\end{minipage}

\caption{
An example with $K = 4$ skill attributes. Left: attribute hierarchy graph $\mce$. Middle: all the allowable attribute patterns in $\mca(\mce)$. Right: $K\times K$ reachability matrix $\mathbf G$.
}
\label{fig-lcbntoy}
\end{figure}
\end{example}
It is worth emphasizing the distinction between a directed acyclic graph (DAG) in the usual attribute hierarchy method and that in a conventional Bayesian network model \citep[][or equivalently, a probabilistic directed graphical model]{pearl1988probabilistic}.
Specifically, the arrows in the DAG among the latent attributes (as shown in Figure \ref{fig-lcbntoy}) generally \textit{cannot} be interpreted as encoding direct statistical dependence, nor does the lack of arrows indicate conditional independence. Rather, such a DAG merely encodes certain hard constraints on what attribute patterns are permissible (those $\aaa\in\mca(\mce)$) and which are forbidden (those $\aaa\in\{0,1\}^K \setminus \mca(\mce)$).
In contrast, the DAG in a Bayesian network has arrows capturing the statistical dependence between the random variables, and the lack of arrows can indicate conditional independence. 
Such a probabilistic DAG generally does not forbid any configurations of the random variables.  

A natural and interesting question is -- can we introduce a new family of models that rigorously unify the above two models and inherit the advantages of both? This question is particularly relevant considering the drawbacks of the existing attribute hierarchy method, including not only the lack of interpretability, but also the lack of statistical parsimony. 
To see this, consider an attribute hierarchy $\mce = \{1\to 2,~ 1\to 3,~ \ldots,~ 1\to K\}$ where the first attribute serves as a common prerequisite for all the $K-1$ remaining attributes. This $\mce$ implies $\mca(\mce) = \{\mathbf{0}_{1\times K},~ (1,\aaa')\text{ for all }\aaa'\in\{0,1\}^{K-1}\}$ with $2^{K-1}+1$ permissible patterns. To model this $\mce$, a conventional attribute hierarchy method would require $2^{K-1}$ free parameters in the latent distribution, because it gives each permissible pattern $\aaa$ an unstructured proportion parameter $p_{\aaa}$. Such a lack of parsimony especially creates statistical and computational challenges when there are a large number of attributes but a limited sample size, as would be the case in fine-grained cognitive diagnosis of many skills in small classroom settings.

\subsection{Latent Conjunctive Bayesian Networks}
This subsection introduces a new family of models for attribute hierarchy in cognitive diagnosis: the Latent Conjunctive Bayesian Networks (LCBNs). 
LCBNs rigorously unify the attribute hierarchy method in educational measurement and the Bayesian network model in statistical machine learning, and inherit the advantages of both.
Our proposal of LCBNs is inspired by another seemingly remote research area -- graphical modeling of genetic mutations in bioinformatics.
Specifically, the conjunctive Bayesian network (CBN) proposed by \cite{beerenwinkel2005learning} and analyzed by \cite{beerenwinkel2007cbn}, models a set of \textit{observed} binary genetic mutations by a partial order, and assign zero probabilities to genotypes (analogue of our skill attribute patterns) that are not compatible with this partial order (analogue of our attribute hierarchy).
An important difference is that, 
genetic mutations are often assumed to be entirely observed without any latent variables \citep{beerenwinkel2005learning, beerenwinkel2006evo, beerenwinkel2007cbn}. In contrast, in our cognitive diagnostic modeling of educational assessment data, the skill attributes are latent constructs that are not directly observable, but rather indirectly measured by item responses. 
We will further discuss the differences between the proposed LCBN and the CBN  in Section \ref{subsec-compare}, after elaborating on their common conjunctive modeling framework for multiple binary random variables. 

We formally define the {latent} conjunctive Bayesian network for the attribute hierarchy.
Introduce $K$ Bernoulli parameters $\bt = (t_1, ..., t_K)^\top \in (0, 1)^K$. For any $k\in[K]$, denote the set of ``parent'' attributes of $\alpha_k$ in the attribute hierarchy graph by $\pa(k)$. The parent attribute of $\alpha_k$ here has the identical definition as the prerequisite attribute of $\alpha_k$.
For example, for the attribute hierarchy shown in Figure \ref{fig-lcbntoy}, $\pa(1)=\pa(2)=\varnothing$ and $\pa(3)=\pa(4) = \{\alpha_1,~\alpha_2\}$.
Now define the  probability mass function of the attribute pattern as follows:
\begin{align}\label{eq:LCBN}
\forall \aaa\in\{0,1\}^K,~~
    p_{\aaa} &= \mathbb{P} (\aaa \mid \bt) = \prod_{k = 1}^{K} \mathbb{P} (\alpha_k \mid \text{pa} (k)), \text{ where } \\
\notag
    \mathbb{P} (\alpha_k \mid \text{pa} (k)) &= {t_k}^{\alpha_k \prod_{\ell=1}^K \alpha_{\ell}^{G_{l, k}}} (1 - t_k)^{(1 - \alpha_k) \prod_{\ell=1}^K \alpha_{\ell}^{G_{l, k}}} \\ \notag
    &= {t_k}^{\alpha_k \prod_{\ell \to k} \alpha_{\ell}} (1 - t_k)^{(1 - \alpha_k) \prod_{\ell \to k } \alpha_{\ell}} 
    \\
    \label{eq-lcbncond}
    &= \begin{cases}
    t_k, & \text{   if } \alpha_k = 1\text{ and } \prod_{\ell \to k} \alpha_{\ell} = 1;\\
    1 - t_k, &  \text{   if } \alpha_k = 0\text{ and }\prod_{\ell \to k} \alpha_{\ell} = 1; \\
    0, & \text{   if } \alpha_k = 1\text{ and }\prod_{\ell \to k} \alpha_{\ell} = 0; \\
    1, &  \text{   if } \alpha_k = 0\text{ and }\prod_{\ell \to k} \alpha_{\ell} = 0.
\end{cases}  
\end{align}
Eq.~\eqref{eq:LCBN} follows the conventional definition of a Bayesian network (i.e., a probabilistic directed graphical model), where the joint distribution of random variables factorizes into the product of conditional distributions of each variable given its parents \citep{bishop2006pattern}.
The conjunctive Bayesian network defined above has an intuitive and natural interpretation. 
This model states that a student can only master attribute $\alpha_k$  if he/she has already mastered every prerequisite attribute for $\alpha_k$; in this case, the mastery of $\alpha_k$ happens with probability 
$$
t_k = \mathbb P(\alpha_k = 1 \mid \alpha_{\ell} = 1 \text{ for all } \ell\in[K] \text{ such that }\ell \to k);
$$
and $1-t_k$ represents the probability of failing to master $\alpha_k$ given the student has already mastered all of its prerequisite attributes.
The last two lines in \eqref{eq-lcbncond} state that, if a student lacks some of $\alpha_k$'s prerequisite skills, then the probability of mastering $\alpha_k$ equals zero and that of not mastering $\alpha_k$ equals one. Therefore, this model exactly respects the usual constraints on permissible/forbidden patterns as a conventional attribute hierarchy method.
One can readily show that the model in \eqref{eq:LCBN}-\eqref{eq-lcbncond} defines a valid joint distribution of attributes. That is, for any $\bo t$, we have $p_{\aaa}=0$ for any $\aaa\not\in\mca(\mce)$ and $\sum_{\aaa\in\{0,1\}^K} p_{\aaa} = \sum_{\aaa\in\mca(\mce)} p_{\aaa} = 1$.

The following example illustrates how the population proportion parameters $\pp=(p_{\aaa})$ are parameterized by CBN parameters $\bo t$.

\begin{example}[Example \ref{ex:toy} continued]\label{ex:toy-continued}

We revisit the attribute hierarchy in Example \ref{ex:toy} and give it an LCBN parametrization.  
By \eqref{eq:LCBN}, the proportion parameters for the permissible attribute patterns in $\mca(\mce)$ in \eqref{eq-ae} can be written as 
\begin{align*}
&p_{0000}=(1-t_1)(1-t_2), \quad
p_{1000}=t_1(1-t_2), \quad
p_{0100}=(1-t_1)t_2,
\\
&p_{1100}=t_1t_2(1-t_3)(1-t_4),\quad
p_{1110}=t_1t_2t_3(1-t_4),\\
&
p_{1101}=t_1t_2(1-t_3)t_4,\quad
p_{1111}=t_1t_2t_3t_4.
\end{align*}
For any $\aaa\not\in\mca(\mce)$, $p_{\aaa}=0$ is naturally guaranteed by following the CBN definition.
Note that if without the CBN assumption, the proportion parameters $p_{\aaa}$ would be only subject to the sparsity constraint $p_{\aaa}=0$ for $\aaa\not\in\mca(\mce)$; in this case, six free parameters would be needed to specify the latent distribution. In contrast, under the CBN, $\aaa$ can be modeled using four Bernoulli parameters: $t_1, t_2, t_3, t_4$. 
In addition to such statistical parsimony, the LCBN model provides intuitive conditional independence statements about the skill attributes. In the current toy example, LCBN asserts that given a student's latent states of the first two basic skills, their mastery of the third and fourth skills are conditionally independent. 
\end{example}

Under our LCBN-based cognitive diagnostic model, the marginal distribution of the observed item response vector of the $i$th student takes the form:
\begin{equation}\label{eq-pmf}
\mathbb{P}(\RR_i =\bo r\mid \TT, \bt, \mce) = \sum_{\aaa \in \{0,1\}^K } 
\underbrace{{t_k}^{\alpha_k \prod_{\ell=1}^K \alpha_{\ell}^{G_{\ell, k}}} (1 - t_k)^{(1 - \alpha_k) \prod_{\ell=1}^K \alpha_{\ell}^{G_{\ell, k}}} }_{p_{\aaa}} \prod_{j = 1}^J \theta_{j, \aaa}^{r_j} (1 - \theta_{j, \aaa})^{1 - r_j},
\end{equation}
for any response pattern $\bo r\in\{0,1\}^J$. The hierarchy $\mce$ implicitly appears in the above distribution through the reachability matrix entries $G_{\ell,k}$. 
The item parameters $\theta_{j,\aaa}$ in \eqref{eq-pmf} are subject to the constraints imposed by the $\QQ$-matrix and can follow various measurement models described in Examples \ref{exp-dina}--\ref{exp-gdina}.
Now we have completed the specification of an LCBN-based cognitive diagnostic model.

\subsection{Comparison of LCBNs and existing models}
\label{subsec-compare}
{We now discuss the difference between our LCBN-based cognitive diagnostic model and the CBN model for genetic mutations proposed by \cite{beerenwinkel2005learning}.
In a CBN, each binary variable $\alpha_k=1$ or $0$ represents a genetic event of whether an amino acid in the genome has mutated or not. There is a partial order (i.e., $\mce$ in our notation) defined on the genetic events such that certain mutations are the prerequisite for others.
Any patient's genetic mutation profile is fully observed as a binary vector $\mathbf X_i = (X_{i1},\ldots,X_{iK})$, and the probability mass function of $\mathbf X_i$ is
\begin{align*}
\mathbb P(\mathbf X_i = \aaa\mid \bo t, \mce) = {t_k}^{\alpha_k \prod_{\ell=1}^K \alpha_{\ell}^{G_{\ell, k}}} (1 - t_k)^{(1 - \alpha_k) \prod_{\ell=1}^K \alpha_{\ell}^{G_{\ell, k}}},\quad \forall \aaa\in\{0,1\}^K.
\end{align*}
In this fully observed CBN model, the hierarchy graph $\mce$ can be directly read off from the set of all the observed binary patterns (\emph{genotypes}) of the patients. In addition, \cite{beerenwinkel2007cbn} showed that the maximum likelihood estimator of parameters $\bo t$ in a CBN actually has a closed-form solution. In contrast, in our LCBN, students' $J$-dimensional item response vectors $\RR_i$ in \eqref{eq-pmf} do not readily reveal the attribute hierarchy graph $\mce$ among the $K$ latent attributes; furthermore, the LCBN parameters $\bo t$ only enter the likelihood through those mixture proportion parameters $p_{\aaa}$ in \eqref{eq-pmf} rather than directly. Therefore, the identifiability issue of LCBNs is nontrivial, and the estimation of the attribute hierarchy and model parameters in LCBNs is not straightforward.

In terms of modeling the binary latent variables,
LCBNs have the advantages of interpretability and statistical parsimony over conventional attribute hierarchy methods and conventional Bayesian networks.
Comparing these two conventional models, the usual attribute hierarchy method has fewer parameters when the hierarchy graph $\mce$ is dense with many arrows, whereas a Bayesian network without the conjunctive assumption (employed by \cite{hu2020bn} for cognitive diagnosis) has fewer parameters when the graph $\mce$ is sparse. 
As concrete examples, consider the three different hierarchies in Figure \ref{fig-divconv} among $K=7$ binary attributes. The numbers of free parameters needed to specify the distribution for the latent $\aaa$ are shown in Table \ref{tab:number of parameters}, from which it is clear that neither a conventional attribute hierarchy method nor a conventional Bayesian network is universally parsimonious. On the other hand, the number of parameters in LCBNs is $K$ for all hierarchies and is universally parsimonious.

\begin{figure}[h!]
\centering
\qquad\qquad
\begin{minipage}[c]{0.31\textwidth}
\resizebox{0.8\textwidth}{!}{
\begin{tikzpicture}[scale=1.2]
	\node (h1)[hiddens] at (1.5, 2) {$\alpha_1$};
    \node (h2)[hiddens] at (0.5, 1) {$\alpha_2$};
    \node (h3)[hiddens] at (2.5, 1) {$\alpha_3$};
    \node (h4)[hiddens] at (0, 0) {$\alpha_4$};
	\node (h5)[hiddens] at (1, 0) {$\alpha_5$};
    \node (h6)[hiddens] at (2, 0) {$\alpha_6$};
    \node (h7)[hiddens] at (3, 0) {$\alpha_7$};
    
    
    \draw[pre] (h1) -- (h2); \draw[pre] (h1) -- (h3);
    \draw[pre] (h2) -- (h4); \draw[pre] (h2) -- (h5);
    \draw[pre] (h3) -- (h6); \draw[pre] (h3) -- (h7);
\end{tikzpicture}
} 
\end{minipage}
\begin{minipage}[c]{0.31\textwidth}
\resizebox{0.8\textwidth}{!}{
\begin{tikzpicture}[scale=1.2]
	\node (h1)[hiddens] at (0, 2) {$\alpha_1$};
    \node (h2)[hiddens] at (1, 2) {$\alpha_2$};
    \node (h3)[hiddens] at (2, 2) {$\alpha_3$};
    \node (h4)[hiddens] at (3, 2) {$\alpha_4$};
	\node (h5)[hiddens] at (0.5, 1) {$\alpha_5$};
    \node (h6)[hiddens] at (2.5, 1) {$\alpha_6$};
    \node (h7)[hiddens] at (1.5, 0) {$\alpha_7$};
    
    
    \draw[pre] (h1) -- (h5); \draw[pre] (h2) -- (h5);
    \draw[pre] (h3) -- (h6); \draw[pre] (h4) -- (h6);
    \draw[pre] (h5) -- (h7); \draw[pre] (h6) -- (h7);
\end{tikzpicture}
} 
\end{minipage}
\begin{minipage}[c]{0.26\textwidth}
\resizebox{0.8\textwidth}{!}{
\begin{tikzpicture}[scale=1.2]
	\node (h1)[hiddens] at (0.5, 2) {$\alpha_1$};
    \node (h2)[hiddens] at (1.5, 2) {$\alpha_2$};
    \node (h3)[hiddens] at (0, 1) {$\alpha_3$};
    \node (h4)[hiddens] at (1, 1) {$\alpha_4$};
	\node (h5)[hiddens] at (2, 1) {$\alpha_5$};
    \node (h6)[hiddens] at (0.5, 0) {$\alpha_6$};
    \node (h7)[hiddens] at (1.5, 0) {$\alpha_7$};
    
    
    \draw[pre] (h1) -- (h3); \draw[pre] (h1) -- (h4);
    \draw[pre] (h1) -- (h5); \draw[pre] (h2) -- (h3);
    \draw[pre] (h2) -- (h4); \draw[pre] (h2) -- (h5);
    \draw[pre] (h3) -- (h7); \draw[pre] (h3) -- (h6);
    \draw[pre] (h4) -- (h7); \draw[pre] (h4) -- (h6);
    \draw[pre] (h5) -- (h7); \draw[pre] (h5) -- (h6);
\end{tikzpicture}
} 
\end{minipage}

\caption{
Different attribute hierarchies with $K=7$ attributes. Divergent (left), convergent (middle), three-layer fully connected (right).
}
\label{fig-divconv}
\end{figure}
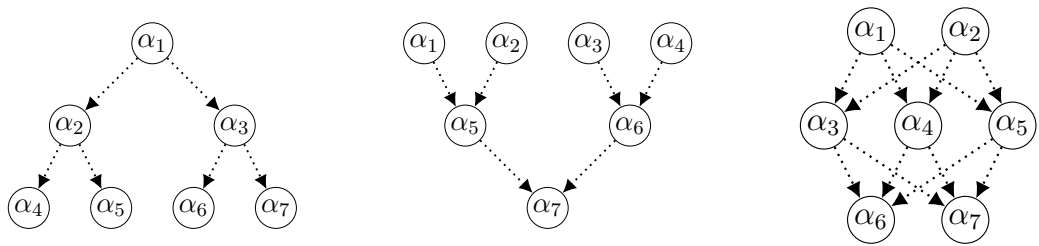

\begin{table}[h!]
\caption{Number of free parameters needed for modeling the latent attributes in the conventional attribute hierarchy method (AHM), Bayesian network (BN), and LCBN with $K = 7$ attributes.}
\label{tab:number of parameters}
\centering
\begin{tabular}{l|ccc}
\hline
Hierarchy $\setminus$ Model & AHM & BN & LCBN \\
\hline
Linear ($\mce=\{1\to 2\to \cdots \to K\}$)      & 7  & 13 & 7 \\
Divergent  (left in Fig. \ref{fig-divconv})      & 25 & 13 & 7 \\
Convergent  (middle in Fig. \ref{fig-divconv})     & 25 & 16 & 7 \\
3-layer fully connected (right in Fig. \ref{fig-divconv})  & 13 & 30 & 7 \\
No hierarchy    & 127& 7 & 7\\
\hline
\end{tabular}

\end{table}

{A related model in the applied psychological measurement literature is the sequential higher order
latent structural model for hierarchical attributes in \cite{zhan2020sequential}.
Specifically, motivated by the higher-order latent trait modeling in \cite{dela2004higher} and the attribute hierarchy method, \cite{zhan2020sequential} proposed a conjunctive model with a higher-order continuous latent variable to model the attributes.
It was assumed that every attribute $\alpha_k$ depends on the higher-order  variable through an item response theory model. 
Our current work differs from this existing work in several fundamental ways. 
First, we do not assume the existence of any higher-order latent variables, which helps achieve the greatest amount of statistical parsimony. Only in this most parsimonious possible LCBN, the lack of arrows between the skills would encode nice conditional independence interpretation; in \cite{zhan2020sequential}'s higher-order model, all the skills are always conditionally dependent due to the higher-order latent trait.
Second, we establish identifiability for the family of LCBN-based cognitive diagnostic models (see Section \ref{sec:id}) and propose a general two-step method to perform both structure learning of $\mce$ and parameter estimation of $(\bo\Theta, \bo t)$ (see Section \ref{sec:est}). In previous studies such as \cite{zhan2020sequential}, identifiability issues were not examined and estimation was performed by assuming the hierarchy $\mce$ is known.
} 

\section{Identifiability of LCBNs for Cognitive Diagnosis}\label{sec:id}

Identifiability is a fundamental property of statistical models and a prerequisite for valid parameter estimation and hypothesis testing. A model is said to be identifiable if the observed data distribution uniquely determines the model parameters. If a model is not identifiable, then there exist multiple and possibly an  infinite number of parameter sets that lead to the same observed distribution, and it is impossible to distinguish them based on data. In the applied context of using LCBNs for cognitive diagnosis, it is crucial to guarantee that the model is identifiable, so that any practical interpretation made about the cognitive structure and student diagnosis is statistically valid.
In this section, we provide transparent conditions for LCBNs to be identifiable.

Because the DINA model in Example \ref{exp-dina} is the most popular and fundamental cognitive diagnostic model due to its interpretability and parsimony, we next focus on the LCBN-based DINA model and provide tight and explicit identifiability conditions for it.
We remark that LCBN-based  CDMs with other measurement models (such as main-effect and all-effect CDMs) are also identifiable under slightly stronger conditions. In light of the space constraint and for notational simplicity, we defer those identifiability results to Section S.1. in the Supplementary Material.

As mentioned earlier, the identifiability of LCBN-based cognitive diagnostic models is a nontrivial and challenging issue, unlike the fully observed CBNs.
Fortunately, thanks to our model formulation, the LCBN parameters $\bo t$ enter the observed distribution in \eqref{eq-pmf} only through the mixture proportion parameters $p_{\aaa}$.
Therefore, we are able to leverage existing techniques for conventional CDMs with an unstructured attribute hierarchy model in \cite{gu2022hlam} to establish identifiability for LCBNs.
Specifically, we next provide conditions that ensure the identifiability of not only the continuous parameters $(\bs, \bg, \bo t)$, but also the discrete hierarchy graph structure $\mce$ in an LCBN.

We first define the concept of strict identifiability of the LCBN-based DINA model.

\begin{definition}[Strict identifiability for LCBN-based DINA]
The parameters $({\mce}, \bs, \bg, {\bt})$ of an LCBN-based DINA model are identifiable if for any $({\mce}, \bs, \bg, {\bt})$ and $(\bar{\mce}, \bar{\bs}, \bar{\bg}, \bar{\bt})$ where $\bar\mce$ induces at most $|\mca(\mce)|$ permisible skill patterns, the following holds if and only if $(\bar{\mce}, \bar{\bs}, \bar{\bg}, \bar{\bt}) = ({\mce}, \bs, \bg, {\bt})$ holds.
\begin{equation}\label{eq:identifiability}
\begin{aligned}
\mathbb{P} (\RR = r \mid \bar{\mce}, \bar{\bs}, \bar{\bg}, \bar{\bt}) =  \mathbb{P} (\RR = r \mid {\mce}, \bs, \bg, {\bt}) \text{ for all } r \in \{0,1\}^J.
\end{aligned}
\end{equation}
\end{definition}

We introduce some new notation before presenting the identifiability results. Following the definition in \cite{gu2022hlam}, 
we categorize the latent attributes into four different types: ancestor, intermediate, leaf, and singleton. An attribute is an ``ancestor attribute'' when it has a child but no parent attribute; an ``intermediate attribute'' when it has both a child and a 
parent; a ``leaf attribute'' when it has a parent but no child attribute; a ``singleton attribute'' when it has no child nor parent attribute. These definitions  are illustrated in Figure \ref{fig-type}.
Interestingly, the identifiability conditions of LCBN-based DINA model can be stated in terms of different types of the attributes in the hierarchy graph.

\begin{figure}[h!]
\centering
\resizebox{0.75\textwidth}{!}{%
\begin{tikzpicture}[scale=1.6]
    \node (h28)[hidden] at (2.6, 2.5) {$\alpha_8$};
    \node (h27)[hidden] at (1.5, 2.5) {$\alpha_7$};
	\node (h26)[hidden] at (0, 2.5)    {$\alpha_6$};
	\node (h25)[hidden] at (-1.1, 2.5) {$\alpha_5$};
    \node (h24)[hidden] at (-2.2, 1.8) {$\alpha_4$}; 
    \node (h23)[hidden] at (-2.2, 3.2) {$\alpha_3$};
    \node (h22)[hidden] at (-3.3, 2.5) {$\alpha_2$};
    \node (h21)[hidden] at (-4.4, 2.5) {$\alpha_1$};
    
    \node (leaftext)[below=0.1cm of h26] {\footnotesize\textbf{Leaf} $\alpha_6$};
    \node [below=0.1cm of h21] {\footnotesize\textbf{Ancestor} $\alpha_1$};
    
    \node [below=0.1cm of h24] {\footnotesize\textbf{Intermediate} $\alpha_2, \alpha_3, \alpha_4, \alpha_5$};
    \node [right=1.5cm of leaftext] {\footnotesize\textbf{Singleton} $\alpha_7, \alpha_8$};
    
    \draw[pre] (h21) -- (h22); 
    \draw[pre] (h22) -- (h23);
    \draw[pre] (h22) -- (h24);
    
    \draw[pre] (h23) -- (h25);
    \draw[pre] (h24) -- (h25);
    \draw[pre] (h25) -- (h26);
\end{tikzpicture}
}
\caption{Illustrating all the four types of attributes in an attribute hierarchy graph.}
\label{fig-type}
\end{figure}
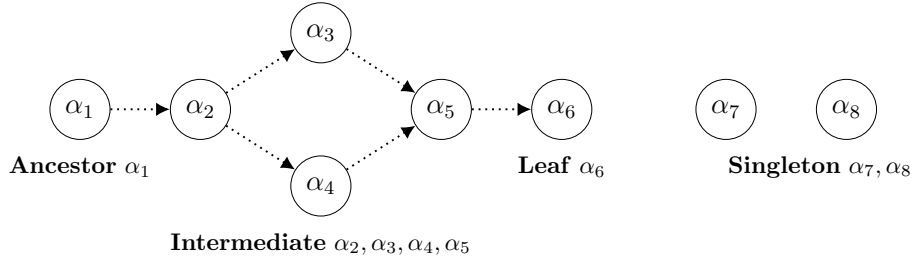

Still following the definition in \cite{gu2022hlam}, we define a ``sparsified'' $\QQ$-matrix given $\mce$ by setting $q_{j,k} = 0$ for any $j, k$ such that $q_{j,h} = 1$ for some child attribute $\alpha_h$ of $\alpha_k$.

\begin{theorem}\label{thm:sid_dina_Q}
The LCBN-based DINA model is strictly identifiable when the $\QQ$ and $\mce$ satisfy the following conditions.

\begin{enumerate}
\item[$A$.] $\QQ$ contains a $K \times K$ submatrix $\QQ_0$ whose sparsified version under $\mce$ is $I_K$. Without the loss of generality, write $\QQ = [\QQ_0^\top, {\QQ^*}^\top ]^\top$.

\item[$B$.] In the sparsified version of $\QQ$, any intermediate attribute is measured at least once, any ancestor or leaf attribute is measured at least twice, and any singleton attribute is measured at least three times.
\item[$C$.] For any singleton attributes $\alpha_k$ and $ \alpha_{\ell}$, the $k$th and $l$th columns of $\QQ^*$ are different.
\end{enumerate}
\end{theorem}

Theorem \ref{thm:sid_dina_Q} is adapted from Theorem 2 in \cite{gu2022hlam} to our LCBN-based model setting.
In general, it is difficult to derive the necessary and sufficient conditions for identifiability of complicated models such as LCBN-based CDMs. Nevertheless, we next show our sufficient identifiability conditions in Theorem \ref{thm:sid_dina_Q} may not be far from being necessary by considering a special hierarchy.
The next proposition states that our conditions $A, B, C$ in  Theorem \ref{thm:sid_dina_Q} become the minimal requirement for identifiability under the linear hierarchy.

\begin{proposition}\label{prop-linear}
Suppose $\mce$ is a linear hiearchy, i.e. $\alpha_1 \rightarrow \alpha_2 \rightarrow \cdots \rightarrow \alpha_K$. Then, the conditions in Theorem \ref{thm:sid_dina_Q} are  necessary and sufficient for strict identifiability of an LCBN-based DINA model. In particular, conditions $B$ and $C$ reduce exactly to be:
\begin{enumerate}
\item[$B^{\star}$.] In the sparsified version of $\QQ$, the leaf attribute and the ancestor attribute are each measured at least twice.
\end{enumerate}
\end{proposition}

The proofs of Theorem \ref{thm:sid_dina_Q} and Proposition \ref{prop-linear}, and additional identifiability results (sufficient conditions for strict and generic identifiability for LCBNs with other measurement models) are included in Sections S.1 and S.2 in the Supplementary Material. 

\section{Two-step Estimation Method for LCBN-based Cognitive Diagnostic Models}\label{sec:est}

This section proposes a two-step estimation method to recover both the attribute hierarchy graph $\mce$ and the model parameters $(\bo\Theta, \bo t)$. Our first step (Algorithm \ref{algo-pem}) uses a penalized EM algorithm under a saturated attribute model to estimate the graph $\mce$, and our second step (Algorithm \ref{algo-em}) develops another EM algorithm to estimate the continuous LCBN parameters. 

We first write out the likelihood given the responses from a sample of $N$ students. 
Denote the response vectors for the $N$ students by
$\RR_i=(R_{i,1},\ldots,R_{i,J})^\top$, for $i=1,\ldots, N$. 
The marginal likelihood under an LCBN-based cognitive diagnostic model is
\begin{align}\label{eq-lk}
L(\TT,\bo t,\mce) 
&= \prod_{i=1}^N
\Big[\sum_{\aaa\in \{0,1\}^K} 
{t_k}^{\alpha_k \prod_{\ell=1}^K \alpha_{\ell}^{G_{\ell, k}}} (1 - t_k)^{(1 - \alpha_k) \prod_{\ell=1}^K \alpha_{\ell}^{G_{\ell, k}}}
\prod_{j=1}^J \theta_{j,\aaa}^{R_{i,j}} (1-\theta_{j,\aaa})^{1-R_{i,j}}  \Big]
\\ \notag
&= \prod_{i=1}^N
\Big[\sum_{\aaa\in \{0,1\}^K} 
p_{\aaa}
\prod_{j=1}^J \theta_{j,\aaa}^{R_{i,j}} (1-\theta_{j,\aaa})^{1-R_{i,j}}  \Big]
=: L(\TT, \pp),
\end{align}
where the last line uses the equivalent parameterization of the mixture proportion parameters $\pp = (p_{\aaa}; \aaa\in\{0,1\}^K)$ instead of the LCBN parameters $\bo t$.
Write the marginal log-likelihood as $\ell(\TT,\bo t,\mce)= \log L(\TT,\bo t,\mce)$ and $\ell(\TT,\pp) = \log L(\TT,\pp)$. 
We next describe the two steps of the proposed estimation procedure in Sections \ref{subsec-1step} and \ref{subsec-2step}, respectively.

\subsection{First step: structure learning of $\mce$ via a penalized EM algorithm}
\label{subsec-1step}

In the first step, we focus on estimating the discrete graph structure in an LCBN: the attribute hierarchy $\mce$. Estimating $\mce$ amounts to performing structure learning of a directed graphical model, and this graphical model is among the $K$ latent skills.

The key idea in learning $\mce$ in an LCBN is to realize that, as an attribute hierarchy $\mce$ naturally defines a set of permissible binary skill patterns $\mca = \mca(\mce)$, a set of permissible patterns $\mca$ also allows for reconstructing an attribute hierarchy graph $\mce = \mce(\mca)$. 
Specifically, one can inversely infer $\mce$ by examining the sparsity structure of $\pp$. 
For a set of permissible patterns $\mca \subseteq \{0,1\}^K$, we can read that $\alpha_k$ is a prerequisite for $\alpha_\ell$ if for any permissible pattern $\aaa = (\alpha_1,\ldots,\alpha_K)\in  \mca$, we have  $\alpha_\ell = 1$ holds only if $\alpha_k = 1$ holds. 
In this way, we can define an attribute hierarchy graph $\mce$ by collecting these prerequisite relationships:
\begin{align}\label{eq-reade}
    \mce = \mce(\mca) = \{k \rightarrow \ell:
    \text{ if for any }\aaa = (\alpha_1,\ldots,\alpha_K)\in  \mca,~~ \alpha_\ell = 1 \text{ only if } \alpha_k = 1
    \}.
\end{align}

\begin{example}[Example \ref{ex:toy} continued]\label{ex-recover}
We revisit the attribute hierarchy $\mce = \{1 \to 3,~ 1 \to 4, ~2 \to 3,~ 2 \to 4 \}$ in Example \ref{ex:toy} and show it can be recovered from the set of permissible patterns. 
First, we collect all the permissible patterns in $\mca$ into a $|\mca| \times K$ matrix denoted by $\mathbf C$. Each row of $\mathbf C$ corresponds to one pattern $\aaa\in\mca$ and each column corresponds to a skill. Then we compare the column vectors of $\mathbf C$ to obtain a partial order among the skills. For example, if $\mathbf C_{:,1} \succeq \mathbf C_{:,3}$ (the first column of $\mathbf C$ is elementwisely greater than or equal to the third column of $\mathbf C$), then it means for all the permissible skill patterns, attribute $\alpha_3$ is present only if attribute $\alpha_1$ is present; this indicates $1 \to 3$. In the current toy example, such a procedure gives the following reconstruction of the hierarchy $\mce$.
\begin{align*}
\mathbf C = 
\begin{pmatrix}
0 & 0 & 0 & 0 \\
1 & 0 & 0 & 0 \\
0 & 1 & 0 & 0 \\
1 & 1 & 0 & 0 \\
1 & 1 & 1 & 0 \\
1 & 1 & 0 & 1 \\
1 & 1 & 1 & 1
\end{pmatrix}
\quad\stackrel{\text{get a partial order between columns}}{\Longrightarrow}\quad 
\begin{matrix}
\mathbf C_{:,1} \succeq \mathbf C_{:,3} \\[2mm]
\mathbf C_{:,1} \succeq \mathbf C_{:,4} \\[2mm]
\mathbf C_{:,2} \succeq \mathbf C_{:,3} \\[2mm]
\mathbf C_{:,2} \succeq \mathbf C_{:,4} \\[2mm]
\end{matrix}
\quad\stackrel{\text{}}{\Longrightarrow}\quad 
\mce = \left\{
\begin{matrix}
1 \to 3,\\
1 \to 4,\\
2 \to 3,\\
2 \to 4.
\end{matrix}
\right\}
\end{align*}
\end{example}

To estimate $\mce$, now the problem boils down to estimating $\mca$. 
To this end, we leverage the log penalty and penalized EM algorithm proposed in \cite{gu2019jmlr} for selecting significant latent patterns.
Consider the truncated $\log$ function 
$$ 
\log_{\rho_N}(p_{\aaa})   =  \log(p_{\aaa}) \cdot  \mathbbm{1}(p_{\aaa}>\rho_N)+  \log(\rho_N)\cdot  \mathbbm{1}(p_{\aaa}\leq \rho_N),
$$
where $\rho_N$ is a small threshold that avoids the singularity issue of the $\log$ function at zero. 
The penalized marginal log likelihood $\ell^{\lambda}(\TT,\pp )$ is defined as
\begin{align}\label{eq-penalized lik}
\ell^{\lambda}(\TT,\pp ) \
= &~ \ell(\TT,\pp ) + \lambda\sum_{\aaa\in\{0,1\}^K} \log_{\rho_N} (p_{\aaa}),
\end{align} 
where $\lambda < 0$ is a tuning parameter controlling the sparsity of $\pp$. 
We maximize $\ell^{\lambda}(\TT,\pp )$ instead of the original marginal log likelihood $\ell(\TT,\pp )$ using the Penalzed EM (PEM) algorithm in \cite{gu2019jmlr}. We restate this algorithm in Algorithm \ref{algo-pem}.
\darkblue{A smaller tuning parameter $\lambda$ (i.e. larger $-\lambda  = |\lambda| > 0$) leads to a stronger penalty and encourages a sparser $\pp$.}

\begin{remark} 
{
One main reason for choosing the log penalty on the proportion parameters $\pp$ over other popular sparsity-inducing penalties is computational convenience.
Among sparsity-inducing penalties,
the $L_0$ penalty is the most direct one that penalizes the number of nonzero entries. Although $L_0$ penalty encourages sparsity and theoretically leads to consistent selection, it is computationally inefficient due to its discontinuous and nonconvex nature \citep{liu2007variable}.
There exist various attempts to replace the $L_0$ penalty with a similar but more tractable objective.
One such example is the popular $L_1$  \citep[Lasso,][]{tibshirani1996regression} penalty. But actually, Lasso turns out to not induce any sparsity on our proportion parameters $\pp$ because
$$\sum_{\aaa \in \{0,1\}^K} |p_{\aaa}| = \sum_{\aaa \in \{0,1\}^K} p_{\aaa} = 1$$
for any $\pp$. Similarly, elastic net regularization \citep{zou2005regularization} also cannot induce sparsity in our setting. 

Compared to the aforementioned penalties, the log penalty proposed by \cite{gu2019jmlr} is preferable as it not only induces nice sparsity on $\pp$, but also allows for efficient and explicit M-step updates for $\pp$ in an EM algorithm.
This follows from the fact that the log penalty can be alternatively viewed as a Dirichlet prior for $\pp$, which is a conjugate prior for the complete data log likelihood:
\begin{align*}
& \ell_{\text{c}} (\TT,\pp \mid \ma)
=  \sum_{\aaa\in\{0,1\}^K} \sum_{i=1}^N \mathbbm{1}(\ma_i=\aaa) \log (p_{\aaa})
\\ \notag 
&\qquad + \sum_{\aaa\in\{0,1\}^K}\sum_{i=1}^N \mathbbm{1}(\ma_i=\aaa) \sum_{j=1}^J \Big[R_{i,j}\log(\theta_{j,\aaa})
 + (1-R_{i,j})\log(1-\theta_{j,\aaa})\Big].
\end{align*}
For more discussions on the connection between the log penalty and the Dirichlet prior in a Bayesian context, please see Remark 12 in \citet{gu2019jmlr}. 
}
\end{remark}

We denote the estimator of the item parameters by $\TT^{{\lambda}}$ and that of the mixture proportion parameters by $\pp^{{\lambda}} = (p^{\lambda}_{\aaa};~ \aaa\in\{0,1\}^K)$. Further, we define the following estimated set of existing skill patterns:
$$
\mca^{\lambda} = \{\aaa \in \{0, 1 \}^K:~ p_{\aaa}^{\lambda}>\rho_N \};
$$
that is, $\mca^{\lambda}$ collects those skill patterns with estimated proportions greater than the threshold $\rho_N$. 
This $\mca^{\lambda}$ is the key quantity that would give an estimate of the attribute hierarchy $\mce^\lambda$.

We consider a sequence of values for $\lambda$ and select an optimal value based on the Extended Bayesian Information Criterion \citep[EBIC,][]{EBIC}:
\begin{equation*}
\text{EBIC}_{\lambda} = -2\ell(\TT^{{\lambda}}, \pp^{{\lambda}}) + (m^\lambda_{\pp} + m_{\TT})\log N + 2\log {2^K - 1 + m_{\TT} \choose m^\lambda_{\pp} + m_{\TT}},
\quad m^\lambda_{\pp} = \lvert \mca^{\lambda}\rvert - 1.
\end{equation*}
In the above display, $m^\lambda_{\pp}$ denotes the number of free parameters in the proportions $\pp^\lambda$ and $m_{\TT}$ denotes the number of free parameters in the item parameters $\TT^\lambda$ (for example, $m_{\TT} = 2J$ for the DINA model, and $m_{\TT} = \sum_{j=1}^J 2^{\sum_{k'=1}^K q_{j, k'}}$ for the GDINA model).
Then we select the optimal tuning parameter $\hat\lambda$ that minimizes the $\text{EBIC}_\lambda$:
$$
\hat{\lambda} = \arg \min_{\lambda} \text{EBIC}_\lambda.
$$

Compared to BIC, EBIC has an additional penalty term for the number of selected parameters and hence favors a more parsimonious model. 
EBIC has been used in related existing works \citep{gu2019jmlr, ma2022learning}, and it also turns out to be especially useful for estimating $\mce$ in LCBNs. In fact, our simulations suggest that the stronger penalty in EBIC is desirable because overselecting non-existing patterns often leads to error in estimating the graph $\mce$, whereas underselecting truly existing patterns can sometimes still suffice for correctly estimating $\mce$ (see Section \ref{sec:sim}). 
\darkblue{
In addition, other popular criteria for model selection such as cross-validation (whose goal is to minimize the prediction error) is not suitable for selecting $\lambda$ here, because it does not take the model sparsity into account. In fact, in preliminary simulations we have found that cross-validation tends to select a larger $\lambda < 0$ with a smaller magnitude than needed, hence resulting in selecting a not sparse enough model. We present such a simulation study in Supplementary Material S.4.2.
}

Finally, given the estimated set of permissible patterns $\mca^{\hat{\lambda}}$, 
we now define our estimate of the attribute hierarchy graph $\mce$ following \eqref{eq-reade}:
\begin{align}\notag
    \hat\mce = \{k \rightarrow \ell:
    \text{ if for any }\aaa = (\alpha_1,\ldots,\alpha_K)\in  \mca^{\hat{\lambda}},~~ \alpha_\ell = 1 \text{ only if } \alpha_k = 1
    \}.
\end{align}

{
Next, we show that our estimator $\hat{\mce}$ is statistically consistent under suitable conditions. We consider the conventional asymptotic setting where the sample size $N$ goes to infinity, but the number of skills $K$ and the number of items $J$ are fixed. Following the assumption in \cite{gu2019jmlr}, we also assume that
the convergence rate of the MLE satisfies
\begin{equation}\label{eq:MLE convergence rate}
    \frac{\ell(\hat{\TT}, \hat{\pp}) - \ell(\hat{\TT}^{\mce}, \hat{\pp}^{\mce})}{N} = O_p(N^{-\delta})
\end{equation}
for some $\delta \in (0, 1]$. Here, $(\hat{\TT}, \hat{\pp})$ is the MLE obtained from maximizing $L(\TT, \pp)$ in \eqref{eq-lk} and $(\hat{\TT}^{\mce}, \hat{\pp}^{\mce})$ is the oracle MLE assuming that the true hierarchy $\mce$ is known. Similar to \cite{gu2019jmlr}, we impose this assumption because the convergence rate of the MLE with an unknown hierarchy (or equivalently, an unknown number of mixture components $|\mca|$) can be slower than the usual parametric rate with $\delta = 1$ \citep{ho2016convergence}.
The following theorem shows the consistency conclusion.

\begin{theorem}\label{thm:hierarchy consistency}
    Consider an identifiable LCBN-based CDM with parameters $(\TT, \bt, \mce)$. Suppose the item parameter $\TT$ satisfies
    \begin{equation}\label{eq:item parameter signal size}
        \theta_{j,\mathbf{1}} - \max_{\aaa \nsucceq \bq_j} \theta_{j,\aaa} \ge c, \quad \forall j \in [J]
    \end{equation}   
    for a constant $c > 0$, and \eqref{eq:MLE convergence rate} holds. Let the threshold be $\rho_N = O(N^{-\delta})$.
    Then, for any sequence $\{\lambda_N \}$ satisfying $\frac{N^{1-\delta}}{|\log \rho_N|} \lesssim - \lambda_N \lesssim \frac{N}{|\log \rho_N|}$,
    we can consistently estimate the attribute hierarchy $\PP(\hat{\mce}^{\lambda_N} = \mce) \rightarrow 1$ as $N \rightarrow \infty$. Here, $\hat{\mce}^{\lambda_N}$ is the estimated hierarchy based on $\mca^{\lambda_N}$.
\end{theorem}

Theorem \ref{thm:hierarchy consistency} also provides theoretical guidelines on choosing the tuning parameters. In particular, we choose $\rho_N = \frac{1}{2N}$ so that it satisfies the condition in Theorem \ref{thm:hierarchy consistency} for any $\delta$.

In addition to the above estimation consistency result for the attribute hierarchy, one could further quantify uncertainty via formal hypothesis testing.
Specifically, we can consider testing the null hypothesis $H_0: \mce = \hat{\mce}$ using additional response data, where $\hat{\mce}$ is the estimated attribute hierarchy. To this end, one may conduct standard goodness of fit tests such as the likelihood ratio test with a $\chi^2$ asymptotic reference distribution.
We leave the detailed development of such hypothesis testing procedures for future research.

}

\begin{algorithm}[h!]
\caption{Penalized EM to learn the attribute hierarchy graph $\mce$ \\ 
(Algorithm 1 in \cite{gu2019jmlr})}
\label{algo-pem}
\SetKwInOut{Input}{Input}
\SetKwInOut{Output}{Output}

\KwData{$\QQ$-matrix $\QQ = (q_{j,k})$, response vectors $(\mathbf R_1^\top, \ldots, \mathbf R_N^\top)^\top$.}

Initialize $\bo\Delta = (\Delta_{\aaa}:\; \aaa\in\{0,1\}^K)$ from the $(2^K-1)$-dimensional probability simplex.

\While{not converged}{
In the $(t+1)$th iteration,
 
\For{$(i,\aaa)\in[N]\times\{0,1\}^K$}{
\begin{align*}
\varphi^{(t+1)}_{i,\aaa}~=~ 
 \frac{\delta^{(t)}_{\aaa}\cdot \exp \Big\{ \sum_{j=1}^J \Big[ R_{i,j}\log(\theta^{(t)}_{j,\aaa}) + (1-R_{i,j})\log(1-\theta^{(t)}_{j,\aaa}) \Big] \Big\} }{\sum_{\aaa'\in \{0,1\}^K} \delta^{(t)}_{\aaa'}\cdot 
 \exp \Big\{ \sum_{j=1}^J \Big[ R_{i,j}\log(\theta^{(t)}_{j,\aaa'}) + (1-R_{i,j})\log(1-\theta^{(t)}_{j,\aaa'}) \Big] \Big\}};
\end{align*}
    }
    \For{$\aaa \in\{0,1\}^K$}{
     $\delta^{(t+1)}_{\aaa} = \max\{c,~\lambda + \sum_{i=1}^N \varphi^{(t+1)}_{i,\aaa}\};$ ~ ($c>0$ is a pre-specified small constant, set to $c=0.01$ throughout the experiments following the suggestion of \cite{gu2019jmlr});\\
    }
    $\pp^{(t+1)}\leftarrow \bo\delta^{(t+1)}/\left(\sum_{\aaa\in \{0,1\}^K}\delta^{(t+1)}_{\aaa}\right);$
        
    \For{$j\in[J]$}{
    $
    \TT^{(t+1)}_j
    = {\arg\max}_{\TT_j} ~ 
    \Big\{\sum_{\aaa}\sum_{i} \varphi^{(t+1)}_{i,\aaa}\sum_{j} \Big[R_{i,j}\log (\theta_{j,\aaa}^{(t)})
 + (1-R_{i,j})\log (1-\theta_{j,\aaa}^{(t)})\Big]\Big\};
    $
    }
  }
  After convergence, use $\mca^{\lambda}, \TT^{{\lambda}}, \pp^{{\lambda}}$ to calculate the EBIC for a sequence of $\lambda < 0$.\\
  Select $\hat{\lambda}$ with the minimum EBIC and recover the hierarchy structure $\mce^{\hat{\lambda}}$ from $\mca^{\hat{\lambda}}$.\\
 \Output{Attribute hierarchy $\mce$.}

\end{algorithm}

\subsection{Second step: parameter estimation of $(\bo\Theta, \pp)$ via another EM algorithm}
\label{subsec-2step}

We next propose another EM algorithm to estimate the continuous LCBN parameters $\bt$ and  $\TT$. 
The previous Algorithm \ref{algo-pem} does not take into account the LCBN structure, but merely focuses on estimating which skill patterns have nonzero proportions in the student population. {Importantly, note that although the hierarchy graph $\hat\mce$ can be read off from the sparsity structure of $\hat\pp^{\lambda}$, the LCBN parameters $\bo t$ \emph{cannot} be read off from the estimated proportion parameters $\pp$. This is because the latter is an overparametrization of the former, and it is not guaranteed that a freely estimated $\pp$ will correspond to a $K$-dimensional LCBN parameters $\bo t=(t_1,\ldots,t_K)$.} 

We next propose another EM algorithm to re-estimate the continuous parameters in LCBN-based cognitive diagnostic models given $\hat\mce$. For each individual $i = 1, ..., N$, denote their latent skill profile by $\ma_i = (A_{i,1}, ..., A_{i, K})$.
We maximize the likelihood in \eqref{eq-lk} with respect to $(\bt, \TT)$ when holding $\mce = \hat\mce$ fixed.
The log likelihood for the complete data $(\ma_i, \RR_i)$, $i=1,\ldots,N$ takes the following form:
\begin{align}\label{eq-complete}
& \ell_{\text{c}}(\TT,\bt \mid \mce)
=  \sum_{\aaa\in\{0,1\}^K} \sum_{i=1}^N \mathbbm{1}(\ma_i=\aaa) 
\prod_{k = 1}^K \left[ {t_k}^{\alpha_k \prod_{\ell=1}^K \alpha_{\ell}^{G_{\ell, k}}} (1 - t_k)^{(1 - \alpha_k) \prod_{\ell=1}^K \alpha_{\ell}^{G_{\ell, k}}} \right]
\\ \notag 
&\qquad + \sum_{\aaa\in\{0,1\}^K}\sum_{i=1}^N \mathbbm{1}(\ma_i=\aaa) \sum_{j=1}^J \Big[R_{i,j}\log(\theta_{j,\aaa})
 + (1-R_{i,j})\log(1-\theta_{j,\aaa})\Big].
\end{align}
Recall that the prerequisite relationships in $\mce$ completely define the reachability matrix entries $G_{\ell, k} = \mathbbm{1}(\ell \to k)$ in the above expression. So the only things that vary in \eqref{eq-complete} are $(\TT,\bt)$.

In the E-step, we evaluate the conditional expectation of \eqref{eq-complete} given the current parameter values $\TT^{(t)}$ and $ \bt^{(t)}$ from the previous iteration. 
It suffices to evaluate the conditional probability of $\mathbbm{1}(\ma_i=\aaa)$, denoted by $\varphi_{i,\aaa} = \mathbb P(\ma_i=\aaa \mid\TT^{(t)},\bt^{(t)})$. See the detailed formula for $\varphi_{i,\aaa}$ in Algorithm \ref{algo-em}.
Then we obtain the following function of $(\TT, \bo t)$:
$$Q(\TT,\bt \mid\TT^{(t)},\bt^{(t)}) =\mathbb E\Big[\ell_{\text{c}}(\TT,\bt \mid \mce) ~\Big\vert~ \mathcal \TT^{(t)},\bt^{(t)}\Big].$$
Next, in the M-step, we seek the maximiziers of the above function and obtain new estimates of the model parameters:
\begin{equation}\label{eq-maxtheta}
(\TT^{(t+1)},\bt^{(t+1)}) = \arg\max_{\TT,\bt} Q(\TT,\bt \mid \TT^{(t)},\bt^{(t)}).
\end{equation}
Every parameter in $(\TT, \bt)$ is continuous, so we set the partial derivative with respect to each of them to zero to seek $(\TT^{(t+1)},\bt^{(t+1)})$. Because $\log t_k$ and $\log(1 - t_k)$ are the only terms in $Q(\TT,\bt \mid\TT^{(t)}, \bt^{(t)})$ that depend on $t_k$, we have a closed-form update of each $t_k$ as follows:
\begin{align*}
    t^{(t+1)}_k = \frac{\sum_{i=1}^N \sum_{\aaa \in\{0,1\}^K} \alpha_{k} \prod_{\ell=1}^K \alpha_{\ell}^{G_{\ell, k}} \varphi^{(t+1)}_{i, \aaa }}{\sum_{i=1}^N \sum_{\aaa \in\{0,1\}^K} \prod_{\ell=1}^K \alpha_{\ell}^{G_{\ell, k}}\varphi^{(t+1)}_{i, \aaa}}.
\end{align*}
As long as we have $\bo t^{(t+1)}$
, we can easily update the mixture proportion parameters $\pp^{(t+1)} = (p^{(t+1)}_{\aaa})$ for the permissible skill patterns by  following the definition in \eqref{eq:LCBN}. 
As for the item parameters $\TT$, we also have closed form updates under some very popular measurement models such as the DINA and GDINA model.
For example, under the DINA model in \eqref{exp-dina} where $\TT$ collects the slipping and guessing parameters $\bs$ and $\bg$, the closed form updates are:
$$
s_j^{(t+1)} = 1 -  \frac{\sum_{i} \sum_{\aaa} R_{i,j} \mathbbm{1}(\aaa\succeq\bq_j) \varphi^{(t+1)}_{i,\aaa}}{\sum_{i} \sum_{\aaa} \mathbbm{1}(\aaa\succeq\bq_j)\varphi^{(t+1)}_{i,\aaa}}, \quad
g_j^{(t+1)} = \frac{\sum_{i} \sum_{\aaa} R_{i,j} \mathbbm{1}(\aaa\nsucceq\bq_j)\varphi^{(t+1)}_{i,\aaa}}{\sum_{i} \sum_{\aaa} \mathbbm{1}(\aaa\nsucceq\bq_j) \varphi^{(t+1)}_{i,\aaa}}.
$$
As for the GDINA model, we present the closed-form parameter updates in Section S.3 of the Supplementary Material.
For LCBNs with certain measurement models such as the main-effect CDMs, there does not exist closed form updates for $\TT$. In this case, one can just perform a gradient-ascent step for $\TT$ that increases $Q(\TT, \bt\mid \TT^{(t)}, \bt^{(t)})$ in \eqref{eq-maxtheta} instead of finding the exact maximizer. Alternatively, one can also apply existing optimization solvers to find an approximate maximizer of $Q(\TT, \bt\mid \TT^{(t)}, \bt^{(t)})$.

\begin{algorithm}[h!]
\caption{EM to estimate LCBN parameters.}
\label{algo-em}
\SetKwInOut{Input}{Input}
\SetKwInOut{Output}{Output}

\KwData{$\QQ$-matrix $\QQ$, response patterns $\{\RR_i: i=1,\ldots,N\}$, attribute hierarchy $\mathcal{E}$. 
}

Initialize $\bt = (t_1, ..., t_{K})$, $\TT$ (subject to the constraints of the $\QQ$-matrix), and $\GG$. 

\While{not converged}{
 In the $(t+1)$th iteration:

\For{$\aaa \in \mca$}{
$$\pp^{(t+1)}_{\aaa} = \prod_{k=1}^K \left(t_k^{(t)}\right)^{\alpha_{k} \prod_\ell\alpha_{\ell}^{G_{\ell,k}}} \left(1 - t_k^{(t)}\right)^{(1 - \alpha_{ k}) \prod_\ell \alpha_{l}^{G_{\ell,k}}};$$
}

        
\For{$(i, \aaa)\in[N]\times \mca(\mce)$}{
\begin{align*}
\varphi^{(t+1)}_{i,\aaa}~=~ 
 \frac{\pp^{(t)}_{\aaa} \cdot \exp \Big\{ \sum_{j=1}^J \Big[ R_{i,j}\log(\theta^{(t)}_{j,\aaa}) + (1-R_{i,j})\log(1-\theta^{(t)}_{j,\aaa}) \Big] \Big\}}{\sum_{\aaa' \in \mca(\mce)} \pp^{(t)}_{\aaa'} \cdot 
 \exp \Big\{ \sum_{j=1}^J \Big[ R_{i,j}\log(\theta^{(t)}_{j,\aaa'}) + (1-R_{i,j})\log(1-\theta^{(t)}_{j,\aaa'}) \Big] \Big\}};
\end{align*}
}
\For{$k \in[K]$}{
\begin{align*}
    t^{(t+1)}_k = \frac{\sum_{i, \aaa} \alpha_{k} \prod_{\ell=1}^K \alpha_{ {\ell} }^{G_{{\ell} , k}} \varphi^{(t+1)}_{i, \aaa }}{\sum_{i, \aaa} \prod_{\ell=1}^K  \alpha_{{\ell} }^{G_{{\ell} , k}}\varphi^{(t+1)}_{i, \aaa }};
\end{align*}
}
   
    \For{$j\in[J]$}{
    $
    \TT^{(t+1)}_j
    = {\arg\max}_{\TT_j} ~ 
    \Big\{\sum_{\aaa}\sum_{i} \varphi^{(t+1)}_{i,\aaa}\sum_{j} \Big[R_{i,j}\log (\theta_{j,\aaa}^{(t)})
 + (1-R_{i,j})\log (1-\theta_{j,\aaa}^{(t)})\Big]\Big\};
    $
    }
  }
After the total $T$ iterations, \\
\Output{ Estimated parameters $\bt,\TT$.}

\end{algorithm}

{
In the following theorem, we show that our two-stage estimation procedure based on Algorithms \ref{algo-pem} and \ref{algo-em} enjoys consistent estimation of not only the attribute hierarchy graph, but also the continuous parameters.

\begin{theorem}\label{thm:parameter consistency}
    Consider an identifiable LCBN-based CDM with parameters $(\TT, \bt, \mce)$, and suppose that the conditions in Theorem \ref{thm:hierarchy consistency} hold. Let $\hat{\mce}$ be the hierarchy estimated from Algorithm \ref{algo-pem}, and let $(\hat{\TT}_N, \hat{\bt}_N)$ be the maximum likelihood estimator of $({\TT}, {\bt})$ given $\hat{\mce}$.
    Then, $(\hat{\TT}_N, \hat{\bt}_N)$ are consistent, i.e. entries in
    $(\hat{\TT}_N,\hat{\bt}_N)$ converge to corresponding entries in $ (\TT,  \bt)$ in probability as $N \to \infty$.
\end{theorem}
}

{
\subsection{Estimation under unknown $\QQ$}
In the previous subsections, we have focused on estimating the LCBN parameters assuming a known and fixed $\QQ$-matrix. This is a common assumption in cognitive diagnostic assessments, because domain experts and test designers often have specified how the test items depend on the latent attributes. 
But sometimes it is of interest to estimate the $\QQ$-matrix directly from data together with other model parameters.
Our two-step estimation procedure 
for LCBNs can be readily extended to such unknown $\QQ$-matrix settings by leveraging existing $\QQ$-matrix estimation methods for traditional CDMs.
We next briefly describe how the exploratory estimation method in \cite{ma2022learning} can be incorporated into our LCBN estimation procedure with an unknown $K$, $\QQ$, and $\mce$.

We briefly sketch the method proposed by \cite{ma2022learning} in Algorithm \ref{algo-ma}. This algorithm includes an additional truncated Lasso penalty \citep[TLP;][]{shen2012likelihood} term on the item parameter matrix $\TT$ to encourage row-wise sparsity. Consequently, the $\QQ$-matrix is recovered by inspecting the sparsity structure of $\TT$. The attribute hierarchy $\mce$ is estimated by comparing the partial orders of the columns of $\TT$, and assigning binary representations to these columns as attribute patterns. 
This Algorithm \ref{algo-ma} can serve as our new first step in the two-step estimation procedure.
Given the estimated $\QQ$-matrix and $\mce$, we can then apply our proposed Algorithm \ref{algo-em} to estimate the continuous LCBN parameters: $\bt$ and $\TT$. 
We present simulation study results in Supplementary Material S.4.4 that demonstrate the good performance of the above estimation method.

\begin{algorithm}[h!]
\caption{Estimate $K$ and discrete structures $\QQ$ and $\mce$ \\ 
(Brief sketch of Algorithms 1 and 2 in  \cite{ma2022learning})}
\label{algo-ma}
\SetKwInOut{Input}{Input}
\SetKwInOut{Output}{Output}

\KwData{Responses $(\mathbf R_1^\top, \ldots, \mathbf R_N^\top)^\top$.}
Set an upper bound for $|\mca|$, the number of latent configurations

\textbf{Step 1:} Use penalized EM assuming sparsity of $\pp$ and $\TT$ to estimate $\TT$ and $|\mca|$\\
\textbf{Step 2:} Construct the $J \times |\mca|$ indicator matrix $\Gamma = \mathbbm{1} (\theta_{j, m} = \max_{l \in [|\mca|]} \theta_{j, l})$\\
\textbf{Step 3:} Plot a DAG based on the partial orders of the columns of $\Gamma$\\
\textbf{Step 4:} Assign binary representations bassed on this DAG and recover $K$ and $\mce$\\
\textbf{Step 5:} Reconstruct each rows of $\QQ$ based on the corresponding row of the $\Gamma$ matrix \\
\Output{Number of latent attributes $K$, Hierarchy structure $\mce$, $\QQ$-matrix $\QQ$}
\end{algorithm}

\section{Simulation Studies}\label{sec:sim}
In this section, we conduct simulation studies under different models and parameter settings to assess the performance of our proposed method. 

\subsection{Parameter estimation of the LCBN-based DINA and GDINA models}
We consider the LCBN-based DINA and GDINA models (see Examples \ref{exp-dina} and \ref{exp-gdina} for the definition of DINA and GDINA models) with $K = 8$ latent attributes and $J = 24$ items. The  $\QQ$-matrix takes the following form:
\begin{equation}\label{eq:example Q}
\QQ = \begin{pmatrix}
\QQ_1\\
\QQ_2\\
\mathbf I_K
\end{pmatrix},\quad \text{where}~~
\QQ_1 = \begin{pmatrix}
1 & 1 & & 0\\
1 & \ddots & \ddots & \\
& \ddots & \ddots & 1 \\
0 & & 1 & 1\\
\end{pmatrix}_{K\times K}
~~\text{and}~~
\QQ_2 = \begin{pmatrix}
1 & 1 & & 0\\
 & \ddots & \ddots & \\
&  & \ddots & 1 \\
0 & &  & 1\\
\end{pmatrix}_{K\times K}.
\end{equation}
Note that $K=8$ is already a relatively large number of attributes in the educational cognitive diagnosis applications.

In all of our simulations, we specify the true hierarchy $\mce$ to be the diamond hierarchy defined in Figure \ref{fig:diamond}. This is a complex multi-layer hierarchy which encodes $|\mca(\mce)| = 15$ permissible patterns. We set the true LCBN parameters as $\bt = (0.9, 0.8, 0.8, 0.7, 0.7, 0.7, 0.6, 0.6)^\top$. Following the definition in \eqref{eq:LCBN}, we can obtain the  mixture proportion parameters for the permissible skill patterns.  All the permissible patterns indexed by $\aaa_1,\ldots,\aaa_{15}$ and 
their corresponding true proportion parameters are presented in Table \ref{tab:diamond}.

We vary the following three aspects in the simulation studies: (1) \textit{measurement model}: DINA and GDINA; (2) \textit{sample size} $N = 500, 1000, 2000$; 
and (3) \textit{noise level of item parameters}. In the LCBN-based DINA model, we use a noise level $r$ to define the slipping and guessing parameters $\bs, \bg$ by
$$s_j = g_j = r \text{  for all  } j = 1,\ldots,J.$$
The larger the noise level $r$ is, the more uncertain one's responses are, and the more challenging it is to estimate the model parameters. Specifically, under DINA, if $r=0$ then there is no uncertainty in one's responses given their latent skills, whereas if $r=0.5$ the responses are purely random noise.
For the LCBN-based GDINA model, we set the positive response probability of the all-zero skill pattern 
to $r$ and that of the all-one skill pattern 
to $1 - r$ (i.e., $\theta_{j, \mathbf 0_K} = r$ and $\theta_{j,\one_K}= 1 - r$); then we define the remaining 
item parameters by setting all main effects and interaction effects of the required attributes in \eqref{eq-gdina}  to be equal.

In each simulation setting, we run 100 independent simulation replications.
We apply our two-step estimation method described in Section \ref{sec:est}. 
The tuning parameter $\lambda$ in Algorithm \ref{algo-pem} (PEM algorithm) is selected from a grid of ten values $\lambda\in\{-0.4,\, -0.8, \,\ldots,\, -3.6,\,-4.0\}$. 
We evaluate the root mean squared errors (RMSE) of the continuous parameters and also the estimation accuracy of the permissible patterns in $\mca$ (this is same as the estimation accuracy of the hierarchy $\mce$).
The RMSE of the proportion parameters $\hat{\pp}$ is computed using the $2^K$--dimensional sparse vector in the probability simplex, i.e. for $C = 100$ simulations,
$$
\text{RMSE} (\hat\pp) = \sqrt{\frac{1}{2^K C} {\sum_{c= 1}^C \sum_{\aaa \in \{0, 1\}^K} (\hat{p}^{(c)}_{\aaa} - p_{\aaa})^2}},
$$
where $\hat{\pp}^{(c)} = (\hat p^{(c)}_{\aaa})$ denotes the estimator from the $c$th simulation replicate.
We sum over all $\aaa \in \{0, 1\}^K$ instead of $\aaa \in \mca$ in order to compute an accurate RMSE even when the estimated $\hat{\mca}$ is incorrect.
The RMSE of the item parameters under DINA is calculated as
$$\text{RMSE} (\hat{\bo\Theta}) 
= 
\sqrt{\frac{1}{2JC} \sum_{c=1}^C \sum_{j=1}^J \left[ (\hat{s_j}^{(c)} - s_j)^2 + (\hat{g_j}^{(c)} - g_j)^2 \right]}.
$$
The RMSEs for the GDINA item parameters and the LCBN parameters $\bt$ are similarly defined. The estimation accuracy of $\mce$ is defined as
$$
\text{Acc}(\hat \mce) = \frac{1}{C} \sum_{c=1}^C \mathbbm{1}(\hat{\mce}^{(c)} = \mce),
$$
where $\hat{\mce}^{(c)} = \mce$ indicates that the entire hierarchy is exactly recovered.
We also compare our final estimated model (denoted by LCBN in the table) to the first-stage estimate (denoted by PEM in the table) by comparing their EBIC values.
The simulation results for the LCBN-based DINA and GDINA are summarized in Tables \ref{tab-dina} and \ref{tab-gdina}, respectively.
The ``argmin EBIC'' column in Table \ref{tab-dina} (or \ref{tab-gdina}) records the percentage of each method (PEM or our two-step procedure) that achieves the minimum EBIC value among the 100 simulation replicates. \darkblue{We report additional simulation details (convergence criteria and the choice of tuning parameters) and computation time in Supplementary Material S.4.1.}

\bigskip
\begin{minipage}{\textwidth}
  \begin{minipage}[c]{0.3\textwidth}
    \centering
    \resizebox{\textwidth}{!}{
    \begin{tikzpicture}[scale=1.4]

    \node (v1)[hidden] at (2, 3) {$\alpha_1$};
    \node (v2)[hidden] at (1, 2) {$\alpha_2$};
    \node (v3)[hidden] at (3, 2) {$\alpha_3$};
    \node (v4)[hidden] at (0, 1) {$\alpha_4$};
    \node (v5)[hidden] at (2, 1) {$\alpha_5$};
    \node (v6)[hidden] at (4, 1) {$\alpha_6$};
    \node (v7)[hidden] at (1, 0)   {$\alpha_7$};
    \node (v8)[hidden] at (3, 0)   {$\alpha_8$};

    \draw[pre] (v1) -- (v2) node [midway,above=-0.12cm,sloped] {}; 
    \draw[pre] (v1) -- (v3) node [midway,above=-0.12cm,sloped] {}; 
    \draw[pre] (v2) -- (v4) node [midway,above=-0.12cm,sloped] {}; 
    \draw[pre] (v2) -- (v5) node [midway,above=-0.12cm,sloped] {}; 
    \draw[pre] (v2) -- (v6) node [midway,above=-0.12cm,sloped] {}; 
    \draw[pre] (v3) -- (v4) node [midway,above=-0.12cm,sloped] {}; 
    \draw[pre] (v3) -- (v5) node [midway,above=-0.12cm,sloped] {};
    \draw[pre] (v3) -- (v6) node [midway,above=-0.12cm,sloped] {}; 
    \draw[pre] (v4) -- (v7) node [midway,above=-0.12cm,sloped] {}; 
    \draw[pre] (v4) -- (v8) node [midway,above=-0.12cm,sloped] {}; 
    \draw[pre] (v5) -- (v7) node [midway,above=-0.12cm,sloped] {};
    \draw[pre] (v5) -- (v8) node [midway,above=-0.12cm,sloped] {}; 
    \draw[pre] (v6) -- (v7) node [midway,above=-0.12cm,sloped] {}; 
    \draw[pre] (v6) -- (v8) node [midway,above=-0.12cm,sloped] {}; 
    
\end{tikzpicture}
}

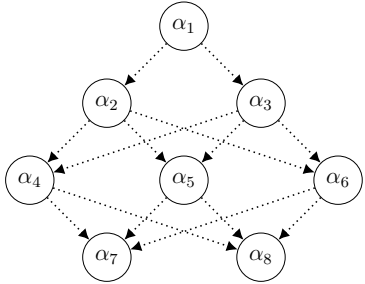
\captionof{figure}{Diamond hierarchy.}
\label{fig:diamond}
\end{minipage}
\hfill
\begin{minipage}[c]{0.65\textwidth}
\centering
\small
\resizebox{0.9\textwidth}{!}{
\begin{tabular}{c|cccccccc|c}
\hline
$\mca(\mce)$  &  $\alpha_1$ & $\alpha_2$ & $\alpha_3$ & $\alpha_4$ & $\alpha_5$ & $\alpha_6$ & $\alpha_7$ & $\alpha_8$ & $p_{\aaa}$\\
\hline
$\aaa_1$ & 0  &   0  &   0    & 0    & 0   &  0   &  0  &   0  & 0.100\\
$\aaa_2$ & 1  &   0  &   0    & 0    & 0   &  0   &  0  &   0  & 0.036\\
$\aaa_3$ & 1  &   0  &   1    & 0    & 0   &  0   &  0  &   0  & 0.144\\
$\aaa_4$ & 1  &   1  &   0    & 0    & 0   &  0   &  0  &   0  & 0.144\\
$\aaa_5$ & 1  &   1  &   1    & 0    & 0   &  0   &  0  &   0  & 0.016\\
$\aaa_6$ & 1  &   1  &   1    & 0    & 0   &  1   &  0  &   0  & 0.036\\
$\aaa_7$ & 1  &   1  &   1    & 0    & 1   &  0   &  0  &   0  & 0.036\\
$\aaa_8$ & 1  &   1  &   1    & 0    & 1   &  1   &  0  &   0  & 0.085\\
$\aaa_9$ & 1  &   1  &   1    & 1    & 0   &  0   &  0  &   0  & 0.036\\
$\aaa_{10}$& 1  &   1  &   1    & 1    & 0   &  1   &  0  &   0  & 0.085\\
$\aaa_{11}$& 1  &   1  &   1    & 1    & 1   &  0   &  0  &   0  & 0.085\\
$\aaa_{12}$& 1  &   1  &   1    & 1    & 1   &  1   &  0  &   0  & 0.032\\
$\aaa_{13}$& 1  &   1  &   1    & 1    & 1   &  1   &  0  &   1  & 0.047\\
$\aaa_{14}$& 1  &   1  &   1    & 1    & 1   &  1   &  1  &   0  & 0.047\\
$\aaa_{15}$& 1  &   1  &   1    & 1    & 1   &  1   &  1  &   1  & 0.071\\
\hline
\end{tabular}
}
\captionof{table}{Permissible patterns under the diamond hierarchy}
\label{tab:diamond}
\end{minipage}
\end{minipage}
\bigskip

Tables \ref{tab-dina} and \ref{tab-gdina} show that our method is effective in recovering the attribute hierarchy $\mce$. 
In particular, the recovery accuracy improves as \darkblue{
    the signal-to-noise ratio increases, i.e. as the sample size $N$ increases and noise level $r$ decreases. In particular, when $N$ is large ($N=2000$), Acc$(\hat{\mce})$ is above 0.97 in all of our simulation settings.
    This observation empirically verifies the identifiability 
    and estimation consistency of $\mce$.
    The accuracy of recovering the hierarchy $\mce$ in Tables \ref{tab-dina} and \ref{tab-gdina} is close to 90\% or higher in all scenarios except for the slightly lower values of 74\% and 52\% when $N = 500$ and $r = 0.2$. These two lower accuracy values correspond to the smallest signal-to-noise settings under DINA and GDINA models. Additionally, the estimation accuracy under GDINA is lower than that under DINA, as it has more item parameters that need to be estimated (in our settings, GDINA has 108 parameters whereas DINA has 48 parameters).
}

\begin{table}[h!]
\caption{Estimation accuracy of attribute hierarchy and RMSE for the estimated parameters for the DINA-based LCBN. The ``argmin EBIC'' column shows the percentage of each method (PEM or proposed) having a smaller EBIC among the 100 simulation replications.}
\label{tab-dina}
\resizebox{\textwidth}{!}{
\centering
\begin{tabular}{cccccccccc}
\toprule
Model & $N$ & $r$ & Method & Acc($\hat{\mce}$) & argmin EBIC & RMSE($\hat{\TT}$) & RMSE($\hat{\pp}$) & RMSE($\hat{\bt}$) \\

\midrule
\multirow{12}{*}{\centering DINA} & \multirow{4}{*}{$500$} &\multirow{2}{*}{$0.1$} 
    &  PEM & -- &  7\%  &  $0.042$  &  $0.005$  &  --  \\
& & & Proposed & 0.92 & 93\% &  $0.029$  &  $0.004$  & $0.042$  \\ 
\cmidrule(lr){3-9}
& &\multirow{2}{*}{$0.2$} 
    &  PEM  & -- &  6\%  & $0.053$  & $0.008$ &  --   \\
& & &  Proposed & 0.74 & 94\% & $0.046$  & $0.006$ &  $0.053$   \\ 
\cmidrule(lr){2-9}
& \multirow{4}{*}{$1000$} &\multirow{2}{*}{$0.1$} 
    &  PEM & -- &  2\%  &  $0.033$  &  $0.004$  &   --  \\
& & & Proposed & 0.98 & 98\% &  $0.021$  &  $0.003$  & $0.027$  \\ 
\cmidrule(lr){3-9}
& &\multirow{2}{*}{$0.2$} 
    &  PEM  & -- &  6\%  & $0.040$  & $0.006$ &  --   \\
& & &  Proposed & 0.94 & 94\% & $0.033$  & $0.004$ &  $0.038$    \\ 
\cmidrule(lr){2-9}
& \multirow{4}{*}{$2000$} &\multirow{2}{*}{$0.1$}
    &  PEM  & -- &  2\%  &  $0.021$ & $0.002$ & --   \\
& & &  Proposed & 0.98 & 98\%  &  $0.015$ & $0.001$ & $0.021$     \\ 
\cmidrule(lr){3-9}
& &\multirow{2}{*}{$0.2$} 
    &  PEM  & -- &  0\%   & $0.029$ & $0.004$ & --   \\
& & &  Proposed & 1.00 &  100\% & $0.021$ & $0.002$ & $0.022$    \\ 

\bottomrule
\end{tabular}
}  
\end{table}

\begin{table}[h!]
\caption{Estimation accuracy of attribute hierarchy and RMSE for the estimated parameters for the GDINA-based LCBN.  The ``argmin EBIC'' column shows the percentage of each method (PEM or proposed) having a smaller EBIC among the 100 simulation replications.}
\label{tab-gdina}
\resizebox{\textwidth}{!}{
\centering
\begin{tabular}{cccccccccc}
\toprule
Model & $N$ & $r$ & Method & Acc($\hat{\mce}$) & argmin EBIC & RMSE($\hat{\TT}$) & RMSE($\hat{\pp}$) & RMSE($\hat{\bt}$) \\

\midrule
\multirow{12}{*}{\centering GDINA} & \multirow{4}{*}{$500$} &\multirow{2}{*}{$0.1$} 
    &  PEM  & --  & 0\% & --      &  0.005  &  --   \\
& & &  Proposed & 0.99 & 100\% & 0.109  &  0.003  &  0.039    \\
\cmidrule(lr){3-9}
& &\multirow{2}{*}{$0.2$} 
    &  PEM  & -- &  5\% & -- &  0.009  &  --   \\
& & &  Proposed & 0.52 &  95\% & 0.176 &  0.004 &  0.086  \\
\cmidrule(lr){2-9}
& \multirow{4}{*}{$1000$} &\multirow{2}{*}{$0.1$}
    &  PEM  & --  & 1\%  & -- &  0.003   &  --  \\
& & & Proposed & 0.99 & 99\%  & 0.072 &  0.002  & 0.030  \\
\cmidrule(lr){3-9}
& &\multirow{2}{*}{$0.2$} 
    &  PEM  & --   & 0\% & --  &  0.007  &  --   \\
& & &  Proposed & 0.89 & 100\% & 0.121  &  0.003  &  0.055   \\
\cmidrule(lr){2-9}
& \multirow{4}{*}{$2000$} &\multirow{2}{*}{$0.1$}
    &  PEM  & --  & 3\%  & -- &  0.002  &  --  \\
& & & Proposed & 0.97 & 97\%  & 0.052 &  0.001  & 0.025  \\
\cmidrule(lr){3-9}
& &\multirow{2}{*}{$0.2$} 
    &  PEM  & --   & 0\% & --  &  0.005  &  --   \\
& & &  Proposed & 0.99 & 100\% & 0.080  &  0.002  &  0.035   \\
\bottomrule
\end{tabular}
}
\end{table}

We also observe that the first-step Algorithm \ref{algo-pem} alone can sometimes under-select the skill patterns when applied to LCBNs. For instance, in our simulations under the diamond hierarchy with 15 permissible patterns, Algorithm \ref{algo-pem} often selects between 11 to 14 patterns without selecting the pattern with the smallest mixture proportion: $\aaa_5 = (1, 1, 1, 0, 0, 0, 0, 0)$ with $p_{\aaa_5} = 0.016$. 
However, this turns out not to be a problem for our two-step LCBN estimation. The reason is that even when some permissible patterns are not detected (i.e., under-selection), it may still be possible to use our method described in Example \ref{ex-recover} to exactly recover the true hierarchy $\mce$.
Indeed, it can be shown that the diamond hierarchy in Figure \ref{fig:diamond} can still be recovered even when as many as five patterns (i.e., $\aaa_2, \aaa_5, \aaa_6, \aaa_{11}, \aaa_{12}$ in Table \ref{tab:diamond}) out of the 15 ones are not detected by the first-step Algorithm \ref{algo-pem}. 

{
In fact, we find in simulations that when the hierarchy is not perfectly estimated, the errors are primarily caused by the over-selection of one additional non-permissible pattern. 
But even in this case, the resulting estimated hierarchy is still close to the truth.
To empirically examine the exact source of uncertainty and inaccuracy, we have performed 200 simulation replications under DINA with $N = 500, r = 0.1$ and inspected the estimated hierarchies.
Out of the 200 replications, 185 ones have exact recovery of the attribute hierarchy. Among the remaining 15 replications, there are at most two prerequisite relations that are not correctly detected.
    In the middle and right panels of Figure \ref{fig:incorrect estimation}, we display examples of such incorrectly estimated hierarchies. In the middle panel, the impermissible skill pattern $(0,0,1,0,0,0,0,0)$ is mistakenly selected, which causes the missingness of the true prerequisite relation $1 \to 3$.
    In the right panel, the impermissible skill pattern $(1,0,0,0,0,1,0,0)$ is mistakenly selected, and hence two prerequisite relations $2 \to 6$ and $3 \to 6$ are missing and an additional arrow $1\to 6$ is detected (note that this arrow is also implied in the true hierarchy).

    The left panel in Figure \ref{fig:incorrect estimation} displays the correct detection percentages for each arrow calculated from the simulation replications. We can see that each prerequisite relation is correctly detected in more than 97\% of the time. 
    These accuracy numbers are larger than  $\text{Acc}(\hat\mce)=92\%$ reported in Table \ref{tab-dina} under the same simulation setting, but this is just because the number 92\% there is calculated as the percentage of times where the entire hierarchy graph $\mce$ is perfectly recovered.
}

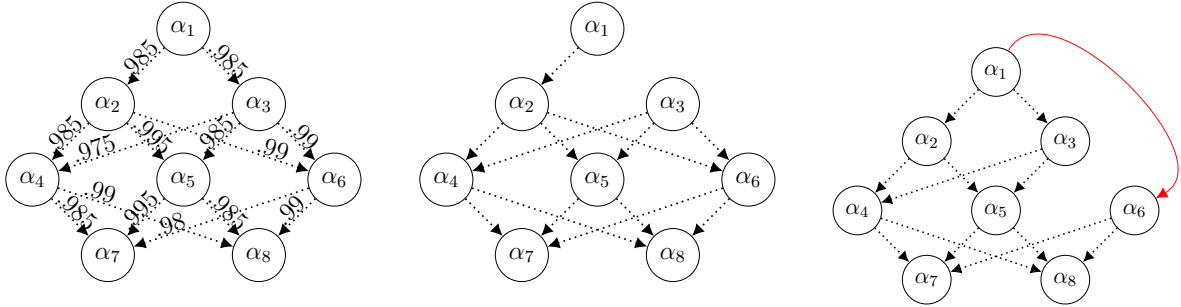
\begin{figure}[h!]
\begin{minipage}{\textwidth}
\centering
  \begin{minipage}[c]{0.3\textwidth}
    \centering
    \resizebox{\textwidth}{!}{
    \begin{tikzpicture}[scale=1.25]

    \node (v1)[hidden] at (2, 3) {$\alpha_1$};
    \node (v2)[hidden] at (1, 2) {$\alpha_2$};
    \node (v3)[hidden] at (3, 2) {$\alpha_3$};
    \node (v4)[hidden] at (0, 1) {$\alpha_4$};
    \node (v5)[hidden] at (2, 1) {$\alpha_5$};
    \node (v6)[hidden] at (4, 1) {$\alpha_6$};
    \node (v7)[hidden] at (1, 0)   {$\alpha_7$};
    \node (v8)[hidden] at (3, 0)   {$\alpha_8$};

    \draw[pre] (v1) -- (v2) node [midway,above=-0.12cm,sloped] {.985}; 
    \draw[pre] (v1) -- (v3) node [midway,above=-0.12cm,sloped] {.985}; 
    \draw[pre] (v2) -- (v4) node [midway,above=-0.12cm,sloped] {.985}; 
    \draw[pre] (v2) -- (v5) node [midway,above=-0.12cm,sloped] {.995}; 
    \draw[pre] (v2) -- (v6) node [midway,above=-0.12cm,sloped] {\hspace{1.7cm}.99}; 
    \draw[pre] (v3) -- (v4) node [midway,above=-0.12cm,sloped] {\hspace{-1.7cm}.975}; 
    \draw[pre] (v3) -- (v5) node [midway,above=-0.12cm,sloped] {.985};
    \draw[pre] (v3) -- (v6) node [midway,above=-0.12cm,sloped] {.99}; 
    \draw[pre] (v4) -- (v7) node [midway,above=-0.12cm,sloped] {.985}; 
    \draw[pre] (v4) -- (v8) node [midway,above=-0.12cm,sloped] {\hspace{-1.7cm}.99}; 
    \draw[pre] (v5) -- (v7) node [midway,above=-0.12cm,sloped] {.995};
    \draw[pre] (v5) -- (v8) node [midway,above=-0.12cm,sloped] {.985}; 
    \draw[pre] (v6) -- (v7) node [midway,above=-0.12cm,sloped] {\hspace{-1.7cm}.98}; 
    \draw[pre] (v6) -- (v8) node [midway,above=-0.12cm,sloped] {.99}; 
\end{tikzpicture}
}
\end{minipage}
\quad
  \begin{minipage}[c]{0.3\textwidth}
    \centering
    \resizebox{\textwidth}{!}{
    \begin{tikzpicture}[scale=1.25]

    \node (v1)[hidden] at (2, 3) {$\alpha_1$};
    \node (v2)[hidden] at (1, 2) {$\alpha_2$};
    \node (v3)[hidden] at (3, 2) {$\alpha_3$};
    \node (v4)[hidden] at (0, 1) {$\alpha_4$};
    \node (v5)[hidden] at (2, 1) {$\alpha_5$};
    \node (v6)[hidden] at (4, 1) {$\alpha_6$};
    \node (v7)[hidden] at (1, 0)   {$\alpha_7$};
    \node (v8)[hidden] at (3, 0)   {$\alpha_8$};

    \draw[pre] (v1) -- (v2) node [midway,above=-0.12cm,sloped] {}; 
    \draw[pre] (v2) -- (v4) node [midway,above=-0.12cm,sloped] {}; 
    \draw[pre] (v2) -- (v5) node [midway,above=-0.12cm,sloped] {}; 
    \draw[pre] (v2) -- (v6) node [midway,above=-0.12cm,sloped] {}; 
    \draw[pre] (v3) -- (v4) node [midway,above=-0.12cm,sloped] {}; 
    \draw[pre] (v3) -- (v5) node [midway,above=-0.12cm,sloped] {};
    \draw[pre] (v3) -- (v6) node [midway,above=-0.12cm,sloped] {}; 
    \draw[pre] (v4) -- (v7) node [midway,above=-0.12cm,sloped] {}; 
    \draw[pre] (v4) -- (v8) node [midway,above=-0.12cm,sloped] {}; 
    \draw[pre] (v5) -- (v7) node [midway,above=-0.12cm,sloped] {};
    \draw[pre] (v5) -- (v8) node [midway,above=-0.12cm,sloped] {}; 
    \draw[pre] (v6) -- (v7) node [midway,above=-0.12cm,sloped] {}; 
    \draw[pre] (v6) -- (v8) node [midway,above=-0.12cm,sloped] {}; 
    
\end{tikzpicture}
}
\end{minipage}
\quad
  \begin{minipage}[c]{0.33\textwidth}
    \centering
    \resizebox{\textwidth}{!}{
    \begin{tikzpicture}[scale=1.25]

    \node (v1)[hidden] at (2, 3) {$\alpha_1$};
    \node (v2)[hidden] at (1, 2) {$\alpha_2$};
    \node (v3)[hidden] at (3, 2) {$\alpha_3$};
    \node (v4)[hidden] at (0, 1) {$\alpha_4$};
    \node (v5)[hidden] at (2, 1) {$\alpha_5$};
    \node (v6)[hidden] at (4, 1) {$\alpha_6$};
    \node (v7)[hidden] at (1, 0)   {$\alpha_7$};
    \node (v8)[hidden] at (3, 0)   {$\alpha_8$};

    \draw[pre] (v1) -- (v2) node [midway,above=-0.12cm,sloped] {}; 
    \draw[pre] (v1) -- (v3) node [midway,above=-0.12cm,sloped] {}; 
    \draw [->,red] (v1) to [out=60,in=30] (v6);
    \draw[pre] (v2) -- (v4) node [midway,above=-0.12cm,sloped] {}; 
    \draw[pre] (v2) -- (v5) node [midway,above=-0.12cm,sloped] {}; 
    \draw[pre] (v3) -- (v4) node [midway,above=-0.12cm,sloped] {}; 
    \draw[pre] (v3) -- (v5) node [midway,above=-0.12cm,sloped] {};
    \draw[pre] (v4) -- (v7) node [midway,above=-0.12cm,sloped] {}; 
    \draw[pre] (v4) -- (v8) node [midway,above=-0.12cm,sloped] {}; 
    \draw[pre] (v5) -- (v7) node [midway,above=-0.12cm,sloped] {};
    \draw[pre] (v5) -- (v8) node [midway,above=-0.12cm,sloped] {}; 
    \draw[pre] (v6) -- (v7) node [midway,above=-0.12cm,sloped] {}; 
    \draw[pre] (v6) -- (v8) node [midway,above=-0.12cm,sloped] {}; 
\end{tikzpicture}
}
\end{minipage}
\captionof{figure}{
\darkblue{
Left: Estimation accuracy for each arrow/prerequisite relationship. Middle and Right: examples of incorrectly estimated hierarchies in the first step. The solid red arrow in the right panel indicates an additional detected arrow which is not in the true $\mce$.
}
}
\label{fig:incorrect estimation}
\end{minipage}
\end{figure}

Tables \ref{tab-dina} and \ref{tab-gdina} also show that the proposed method can accurately estimate the continuous parameters $\pp$, $\TT$, and $\bt$.
Similar to the estimation of $\hat{\mce}$, the estimation error of the continuous parameters is smaller under a smaller noise level $r$, and it decreases as  sample size $N$ increases. This observation again corroborates our identifiability and consistency results of the LCBN model parameters. 
One can also see that the RMSE of $\pp$ and $\TT$ after our second-step algorithm is smaller than the RMSE after just the first-step. This indicates that our second-step estimation procedure improves the overall estimation accuracy, by properly taking into account the LCBN structure. 
{In addition, even when the hierarchy is incorrectly estimated in the first-step, 
the error for estimating the continuous parameters in the second step is still not large. For example, for the $N=500, r=0.1$ row in Table 3, the RMSEs of $\hat{\bo\Theta}$ and $\hat{\bt}$ when the hierarchy is incorrect are 0.031 and 0.103, respectively. These numbers are comparable to the overall average RMSEs of 0.029 and 0.042 in the corresponding row of the table.
}

Finally, the model selected after the second-step tends to have a lower EBIC value compared to the first-step selected model. This demonstrates that our parsimonious LCBN is preferable to the unstructured attribute hierarchy model fitted by PEM.

\subsection{Parameter estimation under misspecified models}
Next, we evaluate our estimation procedure under a misspecified model. 
We still consider $K=8$ attributes and the 15 skill patterns in Table \ref{tab:diamond}. Now instead of using the 15 proportion parameters defined in the last column of Table \ref{tab:diamond}, we generate data using the following vector of proportions for $\aaa_1,\ldots, \aaa_{15}$:
$$\pp = (0.10, 0.04, 0.15, 0.15, \textbf{0}, 0.04, 0.04, 0.09, 0.04, 0.09, 0.09, \textbf{0}, 0.05, 0.05, 0.07)^\top \in \Delta^{15}.$$ 
The above parameter setting is obtained by setting the two smallest entries of $\pp$ in Table \ref{tab:diamond} to zero ($p_{\aaa_5}$ and $p_{\aaa_{12}}$), and renormalizing the other entries to sum up to one. 
Note that this new skill pattern distribution cannot be considered as an LCBN nor as an unstructured attribute hierarchy model under the diamond hierarchy in Figure \ref{fig:diamond}. 
We present the simulation results obtained by still applying the PEM method and our proposed method in Table \ref{tab-dina-mis2}. 
The column Acc$(\hat \mca)$ displays the percentage out of all the simulation replicates where all of the true permissible patterns are successfully selected:
$
\text{Acc}(\hat \mca) = {1}/{C} \sum_{c=1}^C \mathbbm{1}(\mca \subseteq \hat{\mca}^{(c)}).
$

\begin{table}[h!]
\centering
\caption{RMSE for the estimated parameters for the misspecified DINA model. The argmin BIC column shows the percentage of each algorithm having a smaller BIC out of all the simulation replicates.}
\label{tab-dina-mis2}
\begin{tabular}{cccccccc}
\toprule
Model & $N$ & $r$ & Method & Acc($\hat{\mca}$)&  argmin EBIC & argmin BIC & RMSE($\hat{\TT}$)  \\
\midrule 
\multirow{12}{*}{\centering DINA} & \multirow{4}{*}{$500$} &\multirow{2}{*}{$0.1$} 
    &  PEM & - & 4\%  & 11\%  & $0.040$ \\ 
& & & Proposed & 1.00 & 96\%  & 89\%  & $0.028$ \\ 
\cmidrule(lr){3-8}
& &\multirow{2}{*}{$0.2$} 
    &  PEM & - & 18\% & 22\% & $0.050$  \\ 
& & &  Proposed & 0.98 & 82\% & 78\% & $0.045$  \\ 
\cmidrule(lr){2-8} 
& \multirow{4}{*}{$1000$} &\multirow{2}{*}{$0.1$} 
    &  PEM  & - & 26\%  & 45\%  & $0.028$ \\ 
& & &  Proposed & 1.00 & 74\%  & 55\% & $0.023$ \\ 
\cmidrule(lr){3-8} 
& &\multirow{2}{*}{$0.2$} 
    &  PEM  & - & 14\% & 14\% & $0.038$ \\ 
& & &  Proposed & 1.00  & 86\% & 86\% & $0.031$ \\ 
\cmidrule(lr){2-8}
& \multirow{4}{*}{$2000$} &\multirow{2}{*}{$0.2$} 
    &  PEM  & -  & 6\% & 18\% & $0.028$  \\ 
& & &  Proposed & 1.00 & 94\% & 82\% & $0.026$  \\ 
\cmidrule(lr){3-8} 
& &\multirow{2}{*}{$0.3$} 
    &  PEM  & -  & 26\% & 32\% & $0.037$  \\ 
& & &  Proposed & 1.00  & 74\% & 68\% & $0.035$  \\ 

\bottomrule 
\end{tabular}

\end{table}

In the setting of Table \ref{tab-dina-mis-2}, even though the true model does not follow an exact attribute hierarchy in the sense of \eqref{eq:ahm def}, all permissible patterns are correctly selected in almost all simulated settings.
Additionally, even though the true model is not LCBN, our estimation procedure accurately estimates the continuous parameters with similar errors compared to Table \ref{tab-dina}. Also, by comparing our final estimate to the first-stage PEM estimate, it is clear that our second-stage estimation decreases the RMSE in most scenarios.
We also observe that the model selected by the second step generally has a lower EBIC and BIC. This can be explained as the parsimony of LCBNs leads to a more desirable model with a better fit to data.
Notably, this advantage of LCBN is especially apparent in the challenging scenarios where the sample size $N$ is small and the noise level $r$ is large; that is, when we have less information in the data with a small signal-to-noise ratio. In summary, in these small sample and noisy scenarios, adopting our LCBN model by assuming that the latent attributes exhibit certain conditional independence according to the hierarchy graph, not only provides nice practical interpretation, but also improves model fit.

We report some additional simulation results in the Supplementary Material to further support our proposed method. In the Supplementary Material, Section S.4.3 includes simulations when the proportion parameters $\pp$ respect the hierarchy graph but attributes do not exhibit the induced conditional independence asserted by LCBNs; \darkblue{Section S.4.4 includes sensitivity analysis for choosing the tuning parameter $\lambda$ in the log penalty.}

\section{Application to Data from the Trends in Mathematics and Science Study}\label{sec:data}
In this section, we apply the proposed method to analyze an educational assessment dataset from the Trends in Mathematics and Science Study (TIMSS).
TIMSS is a series of international assessments
of fourth and eighth graders' mathematics and science knowledge, involving students in over 60 countries \citep{mullis2012timss}. 
We analyze the TIMSS 2011 Austrian fourth-grade mathematics test data, which is publicly available in the R package CDM \citep{george2016r}. 
The data contains the responses of $N = 4668$ Austrian students to $J = 174$ test items. Educational experts have specified the $K = 9$ fine-grained skill attributes to be: (DA) Data and Applying, (DK) Data and Knowing, (DR) Data and Reasoning, (GA) Geometry and Applying, (GK) Geometry and Knowing, (GR) Geometry and Reasoning, (NA) Numbers and Applying, (NK) Numbers and Knowing, (NR) Numbers and Reasoning
\citep{george2015cognitive}. These nine skill attributes were defined by considering the combinations of three content skills (Data, Geometry, and Number) and three cognitive skills (Applying, Knowing, and Reasoning). This attribute definition follows \cite{george2015cognitive}.
A corresponding $\QQ$-matrix was also specified in \citet{george2015cognitive}. This $\QQ$-matrix assumes that each item measures exactly one attribute, i.e. each row of $\QQ$ is a standard basis vector. 
In this $\QQ$-matrix, each attribute is required by at least six items, so our identifiability conditions in Theorem \ref{thm:sid_dina_Q} are satisfied.

One structure specific to large scale assessments such as TIMSS is that only a subset of all items in the entire study is presented to each of the students \citep{george2015cognitive}. This
results in many missing entries in the $N\times J$ data matrix.
Nevertheless, these entries are missing at random because the missingness patterns do not depend on the students' latent skills or model parameters. Our estimation algorithms can be easily adapted to this setting. Specifically, in the complete data log likelihood used in our EM algorithms, we can just replace the summation range from
$\sum_{i=1}^N \sum_{j=1}^J$ to $\sum_{(i,j) \in \Omega}$, where $\Omega$ is the collection of indices $(i,j)$ that correspond to the observed entries in the data matrix. 

As a first analysis, we apply the two-step  method in Section \ref{sec:est} to estimate the latent hierarchy graph and the continuous parameters.
Note that since the $\QQ$-matrix has all the row vectors being standard basis vectors, the DINA model in Example \ref{exp-dina} and GDINA model
in Example \ref{exp-gdina} are equivalent. So it suffices to just adopt the DINA model in the analysis.
Algorithm \ref{algo-pem} selected 15 attribute patterns, and Figure \ref{fig:hierarchy} shows the reconstructed attribute hierarchy and the estimated latent CBN parameter $\bt = (t_1,\ldots, t_9)$. 
Figure \ref{fig:hierarchy} reveals that there are three ancestor attributes, DK, NK, GK that serve as the prerequisite attributes for each type of content skills in Data, Number, and Geometry. This implies that among the three cognitive skills Knowing, Applying, and Reasoning, the skill Knowing is the most basic. If a student ``Knows'' a certain content skill, then they possesses the prerequisite to ``Apply'' or ``Reason'' the same content skill, sometimes with the aid of other content skills. For instance, NR (Number and Reasoning) requires DR (Data and Reasoning) and DA (Data and Applying) in addition to NA (Number and Applying) as prerequisites.

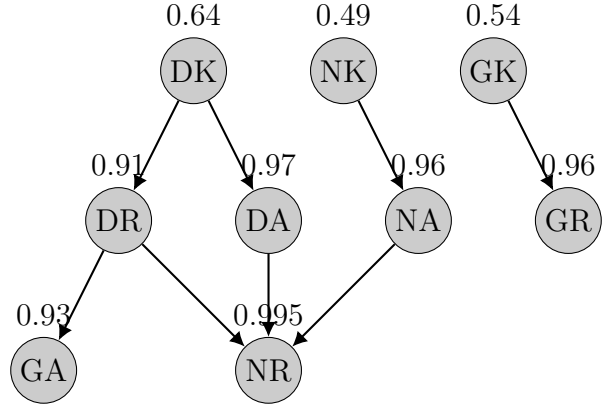
\begin{figure}[h!]
\centering

\vspace{-5mm}
\resizebox{0.5\textwidth}{!}{
    \begin{tikzpicture}[scale=2]

    \node (v1)[neuron] at (1, 2) {DK};
    \node (v2)[neuron] at (2, 2) {NK};
    \node (v3)[neuron] at (3, 2) {GK};
    \node (v4)[neuron] at (0.5, 1) {DR};
    \node (v5)[neuron] at (1.5, 1) {DA};
    \node (v6)[neuron] at (2.5, 1) {NA};
    \node (v7)[neuron] at (3.5, 1) {GR};
    \node (v8)[neuron] at (0, 0)   {GA};
    \node (v9)[neuron] at (1.5, 0) {NR};

    \draw[qedge] (v1) -- (v4) node [midway,above=-0.12cm,sloped] {}; 
    \draw[qedge] (v1) -- (v5) node [midway,above=-0.12cm,sloped] {}; 
    \draw[qedge] (v2) -- (v6) node [midway,above=-0.12cm,sloped] {}; 
    \draw[qedge] (v3) -- (v7) node [midway,above=-0.12cm,sloped] {};
    \draw[qedge] (v4) -- (v8) node [midway,above=-0.12cm,sloped] {}; 
    \draw[qedge] (v4) -- (v9) node [midway,above=-0.12cm,sloped] {}; 
    \draw[qedge] (v5) -- (v9) node [midway,above=-0.12cm,sloped] {}; 
    \draw[qedge] (v6) -- (v9) node [midway,above=-0.12cm,sloped] {}; 
    
    \node[above = 0.02cm of v1] (h) {0.64};
    \node[above = 0.02cm of v2] (h) {0.49};
    \node[above = 0.02cm of v3] (h) {0.54};
    \node[above = 0.02cm of v4] (h) {0.91};
    \node[above = 0.02cm of v5] (h) {0.97};
    \node[above = 0.02cm of v6] (h) {0.96};
    \node[above = 0.02cm of v7] (h) {0.96};
    \node[above = 0.02cm of v8] (h) {0.93};
    \node[above = 0.02cm of v9] (h) {0.995};
    
\end{tikzpicture}
}
\caption{Estimated hierarchy of the TIMSS 2011 dataset. The LCBN parameters $t_k$ are displayed above each skill attribute.
}
\label{fig:hierarchy}
\vspace{-4mm}
\end{figure}

Recall that each $t_k$ gives the conditional probability of mastering attribute $\alpha_k$ provided that one has already mastered all of $\alpha_k$'s prerequisites.
{
One consequence of this definition is that $t_k$ does not capture the individual effect of mastering any specific parent on the mastery of $\alpha_k$. As pointed out by a reviewer, sometimes it may also be interesting to consider such individual effects, e.g., the skill NR in Figure \ref{fig:hierarchy} has three parents and one may wish to distinguish their individual influences.
One possible way to indirectly think about this could be to compare the values of marginal mastery probability $\PP(\alpha_l = 1)$ for each parent skill $l \in \pa(k)$. 
The parent skill $\alpha_l$ with the smallest marginal mastery probability $\PP(\alpha_l = 1)$ could be viewed as having the largest influence on the mastery of the child skill $\alpha_k = 1$.
Going back to the current data example with $k = $NR, the skill NA could be viewed as having the largest influence on NR among the three parent skills since $\mathbb P(\text{NA} = 1)$ is the smallest among those three.
We also include more discussions on potential alternative models to distinguish parent attributes' individual effects in Section \ref{sec:conclusion}.
}

Additionally, Figure \ref{fig:hierarchy} shows that a lot of the $t_k$ parameters in the second or third layer are larger than 0.9, whereas the ancestor attributes have much smaller $t_k$ values. 
Specifically, consider $t_{DA} = 0.97$. Then, $\mathbb P(\alpha_{DA} = 0 \mid \alpha_{DK} = 0) = 1$ and $\mathbb P(\alpha_{DA} = 0 \mid \alpha_{DK} = 1) = 0.03$, whereas
$\mathbb P(\alpha_{DA} = 1 \mid \alpha_{DK} = 1 ) = 0.97$.
This implies that DA may not be a meaningful attribute, as it does not offer additional discrimination of students compared to DK. 
Therefore, we conduct a second analysis and merge those attributes whose $t_k > 0.95$. 
For instance, we combine the attributes ``DA'' and ``DK'' into one ``meta'' attribute. This simplification reduces the number of attributes $K$ from nine to five and the number of permissible attribute patterns $|\mca|$ from 69 to 16. Then we fit an LCBN with this new attribute hierarchy in Figure \ref{fig:hierarchy-2}, where the new $J\times 5$ $\QQ$-matrix can be obtained by summing the corresponding columns in the original $J\times 9$ $\QQ$-matrix. The fitted LCBN parameters are shown in Figure \ref{fig:hierarchy-2}. The final result has the log likelihood equal to $-5.88 \times 10^4$ and EBIC equal to $1.205 \times 10^5$, which is a great improvement compared to the values in our first analysis (previous log likelihood equal to $-6.43 \times 10^4$ and EBIC equal to $1.327 \times 10^5$). 
This implies that merging the attributes and fitting an even more parsimonious LCBN model provides better fit to data. In summary, our LCBN model is a parsimonious and interpretable alternative to existing cognitive diagnostic models, and is especially useful to make sense of data arising from modern large-scale educational assessments such as TIMSS.
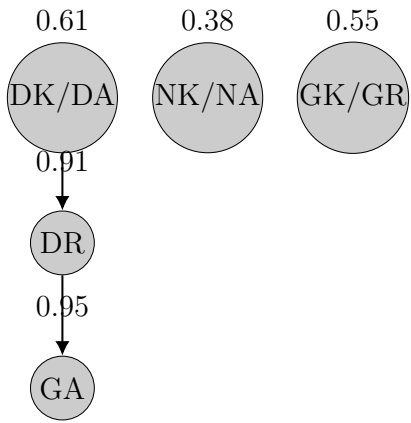
\begin{figure}[h!]
\centering

\vspace{-5mm}
\resizebox{0.34\textwidth}{!}{
    \begin{tikzpicture}[scale=2]

    \node (v1)[neuron] at (1, 2) {DK/DA};
    \node (v2)[neuron] at (2, 2) {NK/NA};
    \node (v3)[neuron] at (3, 2) {GK/GR};
    \node (v4)[neuron] at (1, 1) {DR};
    \node (v5)[neuron] at (1, 0) {GA};

    \draw[qedge] (v1) -- (v4) node [midway,above=-0.12cm,sloped] {}; 
    \draw[qedge] (v4) -- (v5) node [midway,above=-0.12cm,sloped] {}; 
    
    \node[above = 0.02cm of v1] (h) {0.61};
    \node[above = 0.02cm of v2] (h) {0.38};
    \node[above = 0.02cm of v3] (h) {0.55};
    \node[above = 0.4cm of v4] (h) {0.91};
    \node[above = 0.4cm of v5] (h) {0.95};
    
\end{tikzpicture}
}
\caption{Re-estimated hierarchy. The LCBN parameters $t_k$ are displayed above the attributes.
}
\label{fig:hierarchy-2}
\vspace{-4mm}
\end{figure}

\section{Discussion}\label{sec:conclusion}

We have proposed a new family of latent variable models, the latent conjunctive Bayesian networks, for modeling cognitive diagnostic assessment data in education. The LCBN family rigorously unifies the attribute hierarchy method in educational cognitive diagnosis and Bayesian networks in statistical machine learning.
Compared to existing modeling approaches, our model is identifiable, parsimonious, and provides nice interpretation of conditional independence.
We propose a two-step  method that efficiently estimates the discrete attribute hierarchy graph and the continuous model parameters, \darkblue{and establish the consistency of this procedure.
We have also shown that our method can be easily extended to more challenging settings with an unknown $\QQ$-matrix.}
Simulation studies and real data analysis demonstrate that our method has good empirical performance.

Our estimation procedure is scalable and can be easily applied to analyze modern large-scale assessment data,  such as TIMSS and Program for International Student Assessment data. 
Most existing studies of attribute hierarchy focused on the cases when $K = 3$ or 4 due to the computational cost of estimating potentially exponentially many proportion parameters under an unstructured attribute hierarchy model  \citep[e.g.,][]{templin2014hierarchical, wang2021jebs}. On the contrary, our LCBN only requires a linear number of $K$ parameters to specify the latent attribute distribution and  is much more parsimonious. 

\darkblue{
This work proposes the most parsimonious Bayesian network model, LCBN, for attribute hierarchy. In the future, it would also be interesting to explore other Bayesian network models in the cognitive diagnostic applications. 
For example, sometimes the conjunctive assumption in LCBN may be too strong or there may exist multiple paths to master a skill.
To this end, one could consider a latent \emph{disjunctive} Bayesian network:
    \begin{align*}
    \PP(\alpha_k = 1 \mid \alpha_{\pa(k)}) = \begin{cases}
        0,& \text{ if } \prod_{l \in \pa(k)} (1-\alpha_l) = 1,\\
        t_k,& \text{ otherwise}.
    \end{cases}
    \end{align*}
    The above model assumes that as long as a student masters one of the parent attributes of $\alpha_k$, they will have a probability $t_k$ to master $\alpha_k$.
    Alternatively, we could define the following latent \emph{additive} Bayesian network model that defines the conditional mastery probability as a linear combination of those parent attributes:
    \begin{align*}
    P(\alpha_k = 1 \mid \alpha_{\pa(k)}) = \sum_{l \in \pa(k)} t_{k,l} \alpha_l,
    \end{align*}
    where $t_{l,k} \ge 0$ and $\sum_{l \in \pa(k)} t_{l,k} \le 1$. In this model, mastering each parent attribute $\alpha_l$ increases the mastery probability of the child attribute $\alpha_k$ by $t_{k,l}$. This model is less parsimonious than LCBNs, but would be able to model different paths to mastering $\alpha_k$ with different probabilities.
    {It is worth pointing out that the above two alternative Bayesian networks both induce more permissible patterns than the usual attribute hierarchy method described in \cite{leighton2004attribute}. Since the goal of this manuscript is to propose a parsimonious graphical model (i.e., LCBN) for the usual attribute hierarchy, we leave the investigation of  the properties and suitability of the above alternative models for future research.}
}

An interesting future theoretical direction is to study the double-asymptotic regime where $N$ and $J$ both go to infinity and try to consistently estimate the individual-level latent profiles $\mathbf A_i$'s in addition to the model parameters. 
In this work, we study identifiability in the fixed $J$ regime and  focus on identifying and estimating the \emph{population} quantities $(\mce,\bo\Theta, \bt)$.
On the other hand, when $J$ goes to infinity with increasing information provided by each student, 
it may be possible to consistently estimate the individual students' latent skills $\mathbf A_i$ in the \emph{sample} \citep[e.g.,][]{gu2021joint}.
Such sample estimates would provide reliable personalized diagnosis.
Furthermore, if individual students' skills are consistently estimated, then the LCBN parameters can be estimated via a closed form MLE \citep{beerenwinkel2006evo, beerenwinkel2007cbn}.
This can be an alternative estimation method suitable for the double-asymptotic regime without using the regularization as in our current two-step method.
Another interesting future direction is to employ LCBNs in adaptive learning or reinforcement learning settings  \citep{chen2018recommendation, tang2019reinforcement} to help design recommendation strategies and enhance learning. Thanks to LCBNs' parsimony, interpretability, and identifiability, it is attractive to incorporate LCBNs in these computationally intensive applications to help achieve more reliable decision making and recommendations.
We leave these directions for future research.

\spacingset{1}
\bibliographystyle{apalike}
\bibliography{ref}

\renewcommand{\thesection}{S.\arabic{section}}  
\renewcommand{\thetable}{S.\arabic{table}}  
\renewcommand{\thefigure}{S.\arabic{figure}}
\renewcommand{\theequation}{S.\arabic{equation}}
\renewcommand{\thetheorem}{S.\arabic{theorem}}
\renewcommand{\thedefinition}{S.\arabic{definition}}

\setcounter{table}{0}
\setcounter{section}{0}
\setcounter{equation}{0}
\setcounter{figure}{0}
\setcounter{theorem}{0}
\setcounter{definition}{0}
\newpage
\begin{center}
    \LARGE Supplementary Material to ``Latent Conjunctive Bayesian Network: Unify Attribute Hierarchy and Bayesian Network for Cognitive Diagnosis''
\end{center}
\vspace{5mm}

\spacingset{1.45}

This Supplementary Material is organized as follows.
Section \ref{sec:add_id} provides additional identifiability results including conditions for strict and generic identifiability under general LCBN-based CDMs.
Section \ref{sec:proofs} gives proofs for all Theorems in the main paper and Section \ref{sec:add_id}.
Section \ref{sec:gdina update} provides the closed form update for the GDINA item parameters in Algorithm 2 in the main paper.
Section \ref{sec:ad sim} provides many additional simulation results, including: \darkblue{simulation details, a comparison of EBIC and cross validation for choosing $\lambda$, additional simulations under a misspecified model, additional simulations under an unknown $\QQ$-matrix, and sensitivity analysis for choosing the tuning parameter $\lambda$ in the log penalty.}

\section{Additional identifiability results}\label{sec:add_id}
In this section, we provide identifiability results for general LCBN-based CDMs.
First, we define the strict identifiability of LCBNs without assuming a specific measurement model; we state the results in terms of the item parameter matrix $\TT$. Note that assuming a DINA measurement model in the following definition gives Definition 1 in the main paper.
\begin{definition}[Strict identifiability]\label{def:sid}
Consider an LCBN with an attribute hierarchy $\mce$ and model parameters $(\TT, \bt)$. 
For any alternative attribute hierarchy $\bar{\mce}$ which results in at most $|\mca(\mce)|$ permissible patterns, suppose the following inequality holds if and only if $(\mce, \TT, \bt) = (\bar{\mce}, \bar{\TT}, \bar{\bt})$.
\begin{equation}\label{eq-id-def}
\mathbb P(\RR = r \mid \mce, \TT, \bt)
= \mathbb P(\RR = r \mid \bar{\mce}, \bar{\TT}, \bar{\bt}) \text{ for all } r \in \{0,1\}^J
\end{equation}
Then, the parameters $(\mce, \TT, \bt)$ are strictly identifiable.
\end{definition}

Before stating the identifiability result, we need to first introduce some notations.
Recall that $\mca(\mce) \subseteq \{0,1\}^K$ is the set of permissible latent skill patterns that respect an attribute hierarchy $\mce$. 
When it causes no confusion, we also write $\mca(\mce)$ as $\mca$ for notational simplicity.
Similarly to \cite{gu2019jmlr}, we define a binary constraint matrix, $\Gamma^{\mca} \in \{0, 1 \} ^{J \times |\mca|}$, with rows indexed by the $J$ test items and columns by the permissible patterns in $\mca$.
For $j \in [J]$ and $\aaa \in \mca$, the entry 
$\Gamma_{j, \aaa}^{\mca} := \mathbbm{1} (
\aaa \succeq \bo q_j
)$
is a binary indicator of whether pattern $\aaa$ possesses all the required skills of item $j$, because $\bo q_j$ is item $j$'s skill requirement profile. The constraint matrix $\Gamma^{\mca}$ is a function of $\QQ$ and $\mca$.
As a toy example, consider $\mca = \{00, 01, 11\}$ and $\QQ = \mathbf I_2$, then $\Gamma^{\mca}$ takes the form:
$$
\Gamma^{\mca} = 
\begin{blockarray}{ccc}
(00) & (01) & (11)\\
\begin{block}{(ccc)}
0 & 1  & 1\\
0 & 0  & 1\\
\end{block}
\end{blockarray}~.
$$
The $\Gamma^{\mca}$ matrix summarizes the key constraint structure of item parameters $\bo\Theta=(\theta_{j,\aaa})_{J\times|\mca|}$, because Eq.~\eqref{eq-thetaeq}--\eqref{eq-mono} indicate that
\begin{align*}
\theta_{j,\aaa} = \theta_{j,\aaa'} ~\text{ if } ~\Gamma^{\mca}_{j,\aaa} = \Gamma^{\mca}_{j,\aaa'},\quad 
\theta_{j,\aaa} > \theta_{j,\aaa'} ~\text{ if } ~\Gamma^{\mca}_{j,\aaa} > \Gamma^{\mca}_{j,\aaa'}.
\end{align*}
For an item set $S \subseteq [J]$, let $\Gamma^{(S,\mca)}$ denote a submatrix of $\Gamma^{\mca}$ containing the rows indexed by $S$. 
For two skill patterns $\aaa,\aaa'\in\mca$, we write $\aaa\succeq_{S}\aaa'$, if $\Gamma^{\mca}_{j,\aaa}\geq\Gamma^{\mca}_{j,\aaa'}$ for each item $j\in S$. 
This can be interpreted as skill pattern $\aaa$ is at least as capable as $\aaa'$ on the items in the set $S$.
Finally, we say that two item sets $S_1$ and $S_2$ induce the same partial order among the permissible skill patterns, if for any two skill patterns $\aaa$ and $\aaa' \in \mca$, $\aaa\succeq_{S_1}\aaa'$ holds if and only if $\aaa\succeq_{S_2}\aaa'$.
The $\Gamma^{\mca}$ matrix plays an important role in identifiability, as revealed in the following theorem.

\begin{theorem}\label{thm:sid}
A LCBN with a permissible attribute pattern set $\mca=\mca(\mce)$ is strictly identifiable if the binary matrix $\Gamma^{\mca}$ satisfies the following conditions.
\begin{enumerate}
\item[$A^{\star}$.] There exist two disjoint item sets $S_1$, $S_2 \subseteq [J]$, such that $\Gamma^{(S_i,\mca)}$ has distinct column vectors for $i=1,2$; further, $S_1$ and $S_2$ induce the same partial order among the permissible skill patterns in $\mca$.
\item[$B^{\star}$.] For any $\aaa$, $\aaa'\in\mca$ where $\aaa'\succeq_{S_i} \aaa$ under $\Gamma^{\mca}$ for $i=1$ or $2$, there exists some item $j\not\in S_1\cup S_2$ such that $\Gamma_{j,\aaa}^{\mca}\neq \Gamma_{j,\aaa'}^{\mca}$.
\item[$C^{\star}$.] Any column of $\Gamma^{\mca}$ is different from any column of $\Gamma^{\mca^c}$, where $\mca^c = \{0,1\}^K\setminus\mca$.
\end{enumerate}
\end{theorem}

Theorem \ref{thm:sid} is adapted from Theorem 3 in \cite{gu2019jmlr} to our LCBN setting, and guarantees the identifiability of $(\mce, \bo\Theta, \pp)$. Although notation in the theorem may look somewhat heavy, these identifiability conditions are transparent in the sense that they depend only on the binary matrix $\Gamma^{\mca}$, rather than on continuous parameter values of $(\bo\Theta,\pp)$.

Next, we provide sufficient conditions for generic identifiability.
Generic identifiability is a weaker notion compared to strict identifiability, with the intuition that the model parameters and the hierarchy are identified almost surely. The term is formally defined below.

\begin{definition}[generic identifiability]\label{def-gid}
Assume an LCBN with the true hierarchy $\mce$ and parameters $(\TT, \bt)$, where $\TT$ respects the constraints given by $\Gamma^{\mca}$. Denote this constrained parameter space of $(\TT, \bt)$ by $\Omega$. We say $(\mce, \TT, \bt)$ is generically identifiable, if there exists a Lebesgue measure zero subset $\mathcal V \in \Omega$ such that
for any $(\TT,\bt)\in\Omega \setminus \mathcal V$, Equation \eqref{eq-id-def}  
implies $(\mce, \TT, \bt) = (\bar{\mce}, \bar{ \TT}, \bar{\bt})$. 
\end{definition}

The following Theorem \ref{thm-genid} provides sufficient conditions for generic identifiability in terms of the constraint matrix $\Gamma^{\mca}$. These conditions are weaker compared to those in Theorem \ref{thm:sid} and hence are easier to satisfy in practice.

\begin{theorem}\label{thm-genid}
Assume an LCBN with hierarchy $\mce$. If $\Gamma^{\mca}$ satisfies Condition $C$ in Theorem \ref{thm:sid} and also the following conditions, then $(\mce, \TT, \bt)$ is generically identifiable.
\begin{enumerate}
\item[$A^{\star \star}$.] There exist two disjoint item sets $S_1$ and $S_2$, such that altering some entries from 0 to 1 in $\Gamma^{(S_1\cup S_2,\,\mca)}$  can yield a $\widetilde \Gamma^{(S_1\cup S_2,\,\mca)}$ satisfying Condition $A$. That is, $\widetilde\Gamma^{(S_i,\,\mca_0)}$ has distinct columns for $i=1,2$ and $``\succeq_{S_1}" = ``\succeq_{S_2}"$ under $\widetilde\Gamma^{(S_1\cup S_2,\,\mca)}$.

\item[$B^{\star \star}$.] 
For any $\aaa$, $\aaa'\in\mca$ where $\aaa'\succeq_{S_i} \aaa$ under $\widetilde\Gamma^{(S_1\cup S_2,\,\mca)}$ for $i=1$ or $2$, there exists some $j\in(S_1\cup S_2)^c$ such that $\Gamma_{j,\aaa}^{\mca}\neq \Gamma_{j,\aaa'}^{\mca}$.
\end{enumerate}
\end{theorem} 

The proof of Theorem \ref{thm-genid} follows from applying Theorem 2 in \cite{gu2019jmlr}, sharing the spirit of the proof of Theorem \ref{thm:sid}. We omit the details.

Next, we consider the LCBN-based DINA where the $\QQ$-matrix is unknown and also needs to be estimated. The sufficient conditions are provided in Theorem \ref{thm:sid_dina}. We mention that if the $\QQ$-matrix is known, the conditions in Theorem \ref{thm:sid_dina} can be relaxed, as in Theorem \ref{thm:sid_dina_Q}. Indeed, whereas condition B in Theorem \ref{thm:sid_dina} requires every attribute to be measured 3 times, this can be relaxed to being measured twice or once, depending on its type.

\begin{theorem}\label{thm:sid_dina}
The LCBN-based DINA is strictly identifiable upto $\big( \Gamma, \bs, \bg, \bt \big)$when the true (unknown parameter) $\QQ$ and $\mce$ satisfies:
\begin{enumerate}
\item[A1.] $\QQ$ contains a $K \times K$ submatrix $\QQ^0$ that is equivalent to $I_K$ under $\mce$. Without generality, write $\QQ = [\QQ_0^\top, {\QQ^*}^\top ]^\top$.
\item[B1.] The sparsified version of $\QQ$ contains at least three ``1"s in each column.
\item[C1.] The densified version of $\QQ^*$ has distinct columns.
\end{enumerate}
\end{theorem}

The proof of Theorem \ref{thm:sid_dina} follows from applying Theorem 1 in \cite{gu2022hlam}, sharing the spirit of the proof of Theorem \ref{thm:sid_dina_Q}. We omit the details.

 \section{Proof of Theorems \ref{thm:sid}, \ref{thm:sid_dina_Q}, \ref{thm:hierarchy consistency} and Proposition \ref{prop-linear}}\label{sec:proofs}

We next provide the proofs of Theorem \ref{thm:sid}, Theorem  \ref{thm:sid_dina_Q}, and Proposition \ref{prop-linear}, respectively.\\

\noindent
\textbf{Proof of Theorem \ref{thm:sid}}.
We first note that LCBNs can be considered as a structured latent attribute model (SLAM, \cite{gu2019jmlr}) when we reparametrize $\bt, \mce$ as the proportion parameter $\pp$, i.e.
\begin{equation}\label{eq:proportion}
\begin{aligned}
    p_{\aaa} = {{t_k}^{\alpha_k \prod_{\ell=1}^K \alpha_{\ell}^{G_{\ell, k}}} (1 - t_k)^{(1 - \alpha_k) \prod_{\ell=1}^K \alpha_{\ell}^{G_{\ell, k}}} }, \quad \forall \aaa \in \{0, 1 \}^K.
\end{aligned}
\end{equation}
Indeed, equations \eqref{eq-thetaeq} and \eqref{eq-mono} in the main text together with our assumptions for the measurement model, are equivalent to assumptions (2) and (3) in \cite{gu2019jmlr}. Also, our assumptions for the structure model imply $\sum_{\aaa \in \mca} p_{\aaa} = 1$, so the proportion parameter assumption in \cite{gu2019jmlr} is satisfied. It is also easy to check that our three conditions (Conditions $A$, $B$, and $C$) are exactly the same as the identifiability conditions in Theorem 2 in \cite{gu2019jmlr}.

Hence, we can directly apply Corollary 3 in \cite{gu2019jmlr} to obtain that $(\pp, \TT)$ are identifiable. It remains to show that $(\bt, \mce)$ are identifiable from $\pp$. Suppose that $\pp$ is the attribute proportion generated from an LCBN with true parameters $(\bt, \mce)$. Let 
$$\mca = \{\aaa \in \{0, 1 \}^K : p_{\aaa} > 0 \}$$ 
(this is the true $\mca$ by definition).
Then, $\mce$ can be identified by defining $\mce$ based on \eqref{eq-reade}.
Finally, given $\mca$ and $\mce$, $\bt$ can be identified by solving \eqref{eq:proportion} for all $\aaa \in \mca$. Note that under any hierarchy $\mce$,  the number of permissible patterns is at least as many as $K$ (i.e., $|\mca(\mce)| \ge K$) and that \eqref{eq:proportion} contains at least $K$ linear independent constraints. In addition, we know that there exists a true LCBN parameter vector $\bt$ that satisfies \eqref{eq:proportion}, so this is the unique solution to \eqref{eq:proportion}. This proves $\bt$ are also identifiable and completes the proof of the theorem.
\qed
\\

\textbf{Proof of Theorem \ref{thm:sid_dina_Q}.}
Similar to the proof of Theorem \ref{thm:sid}, note that LCBNs can be considered as a hierarchical latent attribute model (HLAM, \cite{gu2022hlam}) when we reparametrize $\bt, \mce$ as the proportion parameter $\pp$ using \eqref{eq:proportion}.
It is also easy to check that our three conditions (Conditions $A$, $B$, and $C$) are exactly the same as the conditions in Theorem 2 in \cite{gu2022hlam}. Hence, we can apply Theorem 2 in \cite{gu2022hlam} to obtain that $(\bs, \bg, \mce, \pp)$ is identifiable. Then similarly to the proof of Theorem \ref{thm:sid}, we again identify $\bt$ from $\pp$ by using \eqref{eq:proportion}. This completes the proof of Theorem \ref{thm:sid_dina_Q}.
\qed
\\

\textbf{Proof of Proposition \ref{prop-linear}}
The sufficiency follows from Theorem \ref{thm:sid_dina_Q}. We prove that both conditions $A$ and $B^{\star}$ are necessary. In this proof, we use the following equivalent parametrization of the slipping and guessing parameters $\bs, \bg$ for notational simplicity.
$$
\theta^+_j = 1 - s_j, \quad\theta^-_j = g_j, \quad \forall j \in [J].
$$

\textit{Necessity of Condition $A$.}

Under the linear hierarchy, the proof of Proposition 3 in \cite{gu2022hlam} can be applied directly. Suppose the sparsified $\QQ$-matrix does not contain $e_h$ for some $1 \le h \le K$. Then, $\aaa_1 = (1, \cdots, 1, 0, 0, \cdots, 0)$ and $\aaa_2 = (1, 1, \cdots, 1, 1, 0, \cdots, 0)$ are configurations in $\mca(\mce)$ with the same ideal response vector across all the items $\Gamma^{\mca}_{:,\aaa_1} = \Gamma^{\mca}_{:,\aaa_2}$ (here, $\aaa_1$ and $\aaa_2$ only differs in the $h$th entry). Hence, $p_{\aaa_1}$ and $p_{\aaa_2}$ are only identifiable up to their sum, so the LCBN parameters $t_h$ and $t_{h+1}$ are identifiable only up to their product.
\medskip

\textit{Necessity of Condition $B^{\star}$.}


(i) We prove that any ancestor attribute needs to be measured at least twice for identifiability to hold. Suppose some ancestor attribute is measured only once, and assume that attribute 1 is an ancestor attribute measured only by item 1. Denote the true LCBN parameters by $(\theta^+, \theta^-, \mce, \bt)$ and the true proportion parameters by $\pp$. We show that there exists $(\bar{\theta}^+, \bar{\theta}^-, \bar{\mce}, \bar{\bt}) \neq (\theta^+, \theta^-, \mce, \bt)$ with the same marginal distributions. 

Define $\bar{\theta}_j^+ = \theta_j^+$ for all $j$, $\bar{\theta}_j^- = \theta_j^-$ for $j \ge 2$, $\bar{\mce} = \mce$, $\bar{t}_k = t_k$ for $k \ge 3$. There are three free parameters: $\bar{\theta_1}^-, \bar{t}_1, \bar{t}_2$.
By the proof of Proposition 5 in \cite{gu2022hlam}, the marginal distribution 
 of the response vector $\mathbf R$ is the same under the true and alternative parameters if the following equations hold:
\begin{align*}
\begin{cases}
    \bar{p}_{(0,\textbf{0}_{K-1})} + \bar{p}_{(1,\textbf{0}_{K-1})} = p_{(0,\textbf{0}_{K-1})} + p_{(1,\textbf{0}_{K-1})} \\
    \bar{\theta}_1^- \bar{p}_{(0,\textbf{0}_{K-1})} + \theta_1^+ \bar{p}_{(1,\textbf{0}_{K-1})} = \theta_1^- p_{(0,\textbf{0}_{K-1})} + \theta_1^+ p_{(1,\textbf{0}_{K-1})}
\end{cases}
\end{align*}
Writing the above equations in terms of $\bar{t}_1$ and $\bar{t}_2$ gives
\begin{align*}
\begin{cases}
    \bar{t}_1 \bar{t}_2 = t_1 t_2 \\
    \bar{\theta}_1^- (1 - \bar{t}_1) + \theta_1^+ \bar{t}_1(1 - \bar{t}_2) = \theta_1^- (1 - {t_1}) + \theta_1^+ t_1(1 - t_2).
\end{cases}
\end{align*}
There are three variables $(\bar t_1, \bar t_2, \bar \theta_1^-)$ that need to satisfy two equations, so there are infinitely many solutions. Hence, the model is not identifiable. This proves that the condition that any ancestor attribute needs to be measured at least twice is necessary for identifiability.
\qed

\begin{remark}
Suppose condition $A$ holds. Then, the above proof only uses the assumption that ``$\alpha_1$ is the only ancestor / singleton attribute''.
\end{remark}

(ii) We prove that $\alpha_K$ needs to be measured at least twice. Suppose not, and assume that attribute $K$ is a leaf attribute measured only by item $K$. We next show that there exists $(\bar{\theta^+}, \bar{\theta^-}, \bar{\mce}, \bar{\bt}) \neq (\theta^+, \theta^-, \mce, \bt)$ that lead to the same marginal distributions for the observed response vector $\mathbf R$. 

Define $\bar{\theta_j^+} = \theta_j^+$ for $j \neq K$, $\bar{\theta_j^-} = \theta_j^-$ for all $j$, $\bar{\mce} = \mce$, $\bar{t_k} = t_k$ for $k \neq K-1, K$. There are three free parameters: $\theta_K^+, t_{K-1}, t_K$.
Similar to the previous argument, the marginal distribution of $\mathbf R$ is the same if the equations
\begin{equation}\label{eq:leaf}
\begin{aligned}
\begin{cases}
    \bar{p}_{(\aaa', 0)} + \bar{p}_{(\aaa', 1)} = p_{(\aaa',0)} + p_{(\aaa',1)} \\
    \theta_K^- \bar{p}_{(\aaa', 0)} + \bar{\theta}_K^+ \bar{p}_{(\aaa', 1)} = \theta_K^- p_{(\aaa', 0)} + \theta_K^+ p_{(\aaa', 1)}
\end{cases}
\end{aligned}
\end{equation}
hold for all $\aaa' \in \{0, 1\}^{K-1}$. Now, note that our parameter assumptions give that $\bar{p}_{(\aaa',0)} = p_{(\aaa',0)}$ and $\bar{p}_{(\aaa',1)} = p_{(\aaa',1)}$ for all $\alpha' \neq \textbf{1}_{K-1}$. Hence, \eqref{eq:leaf} is automatically satisfied except when $\alpha' = \textbf{1}_{K-1}$. Writing this in terms of $\theta_K^+, t_{K-1}, t_K$, \eqref{eq:leaf} is equivalent to
\begin{align*}
\begin{cases}
    \bar{t}_{K-1} = t_{K-1} \\
    {\theta_K^-} (1 - \bar{t}_K) + \bar{\theta}_K^+ \bar{t}_K = \theta_K^- (1 - {t_K}) + \theta_K^+ t_K.
\end{cases}
\end{align*}
Clearly there are infinitely many solutions and this model is not identifiable.
This proves that the condition that any leaf attribute needs to be measured at least twice is necessary for identifiability.

\begin{remark}
Suppose condition $A$ holds. Then, the above proof only uses the assumption that ``$\alpha_K$ is the only leaf / singleton attribute''.
\end{remark}

{
\textbf{Proof of Theorem \ref{thm:hierarchy consistency}.}
    Our proof is mainly based on Theorem 13 in \cite{gu2019jmlr} (denoted as Theorem 13 in GX for simplicity). We first check their conditions. Similar to the proof of Theorem \ref{thm:sid}, we note that LCBN-based CDMs can be viewed as a structured latent attribute model (SLAM) with parameters $(\TT, \pp)$ with $\pp$ given by the LCBN parametrization in \eqref{eq:proportion}. Moreover, as we have that our LCBN-based CDM is identifiable, the corresponding SLAM is also identifiable. The first line in Theorem 13 in GX is only used to guarantee model identifiability, and can be replaced by our assumption.
    
    Noting that we are considering a fixed $K$ and true parameters $t_k \in (0, 1)$, there exists a constant $c_0 > 0$ such that $p_{\aaa} > c_0$ for all $\aaa \in \mca$. Combining this with \eqref{eq:item parameter signal size}, equation (20) in Theorem 13 in GX holds. Finally, as we consider $\mca_{\text{input}} = \{0, 1\}^K$, $|\mca_{\text{input}}| = 2^K$ is a constant with respect to $N$. Hence, every assumption in Theorem 13 in GX holds and we get
    $$\PP(\hat{\mca}^{\lambda_N} = \mca) \rightarrow 1$$
    as $N \rightarrow \infty$.
    Now note that the hierarchy $\mce$ is correctly estimated when the set of permissible patterns $\mca$ is correctly estimated. So we have
    \begin{align*}
        \PP(\hat{\mce}^{\lambda_N} = \mce) \ge \PP(\hat{\mca}^{\lambda_N} = \mca) \rightarrow 1
    \end{align*}
    and $\hat{\mce}^{\lambda_N}$ is consistent.
\qed
\\
}

{
\textbf{Proof of Theorem \ref{thm:parameter consistency}.}
For any vector $\bo a$, let $\|\bo a\|$ denote its $L_2$ norm.
With a slight abuse of notation, let $\TT$ also denote the vector (in addition to its original definition of being a matrix) collecting all the different item parameters in matrix $\TT$, and let $\|\TT\|$ denote the $L_2$ norm of this long item parameter vector.
Let $\hat\mce =: \hat\mce_N$ denote the estimator of the attribute hierarchy graph when sample size is $N$.
For any $\epsilon>0$,
    \begin{align*}
        & \limsup_{N \rightarrow \infty} \PP \left(\|\hat{\TT}_N - \TT \| > \epsilon,~ \|\hat{\bt}_N-\bt\| > \epsilon \right) \\
        =& \limsup_{N \rightarrow \infty} \Big(\PP \left(\|\hat{\TT}_N - \TT \| > \epsilon,~ \|\hat{\bt}_N-\bt\| > \epsilon,~ \hat{\mce}_N = \mce \right) + \\
        & \qquad \qquad \PP \left( \|\hat{\TT}_N - \TT \| > \epsilon,~ \|\hat{\bt}_N - \bt\| > \epsilon,~ \hat{\mce}_N \neq \mce \right) \Big)\\
        \le & \limsup_{N \rightarrow \infty} \left(\PP \left(\|\hat{\TT}_N - \TT \| > \epsilon,~ \|\hat{\bt}_N-\bt\| > \epsilon \mid \hat{\mce}_N = \mce \right) \PP(\hat{\mce}_N = \mce) + \PP(\hat{\mce}_N \neq \mce)\right) \\
        \le & \limsup_{N \rightarrow \infty} \PP \left(\|\hat{\TT}_N - \TT \| > \epsilon,~ \|\hat{\bt}_N-\bt\| > \epsilon \mid \hat{\mce}_N = \mce \right).
    \end{align*}
    The last inequality follows from the fact that we assume the conditions in Theorem \ref{thm:hierarchy consistency}, so we have $\PP(\hat{\mce} = \mce) \rightarrow 1$ and $\PP(\hat{\mce} \neq \mce) \rightarrow 0$.
    Hence, it suffices to show that the last line in the above display is zero for any $\epsilon>0$.
    That is, we only need to show that the MLE $(\hat{\TT}_N, \hat{\bt}_N)$ given the true hierarchy $\mce$ is consistent. This is true following from a standard textbook argument, e.g. Theorem 10.1.6 in \cite{casella2021statistical}, because the continuous parameters are identifiable, and the likelihood given the hierarchy is differentiable. So $ \lim_{N \to \infty} \PP \left(\|\hat{\TT}_N - \TT \| > \epsilon,~ \|\hat{\bt}_N - \bt\| > \epsilon \right) \to 0$, which proves the consistency.
    \qed
}

\section{Closed-form updates for the GDINA model parameters in Algorithm \ref{algo-pem}}\label{sec:gdina update}
Let $S_j = \{k \in [K]: q_{j,k} = 1 \}$. By the $\QQ$-matrix constraints in \eqref{eq-thetaeq} and \eqref{eq-mono}, we can assume without loss of generality that $\delta_{j, S}$ is 0 for $S \not \subseteq S_j$. Hence, it suffices to update the parameters where $S \subseteq S_j$, which can be written as:
\begin{align*}
    \delta_{j, S}^{(t+1)} = \frac{\sum_i  \sum_{\aaa} \mathbbm{1}(\{k \in S_j : \alpha_k = 1\} = S) R_{i,j}  \varphi_{i, \aaa}^{(t+1)}}
    {\sum_i \sum_{\aaa} \mathbbm{1}(\{k \in S_j : \alpha_k = 1\} = S) \varphi_{i, \aaa}^{(t+1)}},\quad \forall j\in[J],~~ S \subseteq S_j.
\end{align*}
The above updates can be used in the M step of Algorithm \ref{algo-pem} for the GDINA model.

\section{Additional simulation studies and details}\label{sec:ad sim}

{
\subsection{Additional simulation details}
For the implementation of our EM algorithms, we made the following specifications. First, we set the convergence criterion of EM Algorithms \ref{algo-pem} and \ref{algo-em} to be that, the algorithm is terminated when the increment of the log likelihood in two consecutive iterations is less than $0.05, 0.01$, respectively. The threshold is larger for Algorithm \ref{algo-pem} as it only aims to estimate the discrete structure, and does not need to estimate continuous parameters very accurately. Note that our convergence criterion is already very stringent when considering the magnitude of the actual likelihood, which is of a $10^4$ scale in both the simulation studies and the real data analysis. The other threshold values in Algorithm \ref{algo-pem} are set to be $c = 0.01$ and $\rho_N = \frac{1}{2N}$ following the suggestions of \cite{gu2019jmlr}.

Next, we report the average number of iterations and runtime for Algorithms \ref{algo-pem} and \ref{algo-em} in Table \ref{tab:time}. The reported values are averages from 100 independent simulation replicates.
Table \ref{tab:time} shows that our new algorithm for LCBN takes fewer than 10 iterations and 25 seconds on average to reach convergence, and the overall two-step estimation procedure is also computationally quite efficient.
We can see that the absolute value of the selected $\hat{\lambda}$ increases with respect to $N$, which is consistent with the asymptotic conditions for $\lambda$ in Theorem \ref{thm:hierarchy consistency}.
Here, we consider the DINA measurement model with noise level $r = 0.1$, and use the same diamond hierarchy and $\QQ$-matrix as in Section \ref{sec:sim} (in other words, Table \ref{tab:time} corresponds to the rows with $r = 0.1$ in Table \ref{tab-dina} in the main body of the manuscript).

Finally, all simulations in this paper were performed in MATLAB on a personal laptop with GPU: Intel Iris Xe graphics card, CPU: Intel i7-1260P Processor with vPro (16GB).

\begin{table}[h!]
\centering
\caption{The average number of iterations and runtime for the EM algorithms}
\label{tab:time}
\begin{tabular}{cccccc}
\toprule
$N$ & Algorithm & \# of iterations & \# of iterations for $\hat{\lambda}$ & runtime (s) & $\hat{\lambda}$  \\
\midrule 
\multirow{3}{*}{$500$} 
& PEM & 66.2 & 16.3 & 11.1 & -3.1 \\
& LCBN & 9.4 & -    & 6.0  & - \\
\cmidrule{2-6}
& Total & 75.6 & - & 17.1 & - \\
\midrule
\multirow{3}{*}{$1000$} 
& PEM & 81.0 & 21.8 & 26.6 & -3.2 \\
& LCBN & 9.0  & -    & 9.7 & - \\
\cmidrule{2-6}
& Total & 90.0 & - & 36.3 & - \\
\midrule
\multirow{3}{*}{$2000$} 
& PEM & 101.5 & 18.2 & 67.0 & -3.4 \\
& LCBN & 9.9  & -    & 22.1 & - \\
\cmidrule{2-6}
& Total & 111.4 & -  & 89.1 & - \\
\bottomrule 
\end{tabular}
\end{table}
}

\subsection{Comparison of EBIC and cross validation for choosing $\lambda$}

{
Next, we compare the EBIC and cross validation (CV) in terms of selecting the tuning parameter 
$\lambda$. We work on the same setting as Section 5.2 with sample size $N = 500$, noise level $r = 0.2$, and $\lambda \in \{0, -0.3, \ldots, -6\}$.

We have tried selecting $\lambda$ via the 5-fold CV (we divide the $N$ subjects into five equal-sized folds), and found that CV tends to select a large $\lambda < 0$ (i.e., small $|\lambda|$) compared to the $\lambda$ chosen via EBIC. This means CV favors a small magnitude of the penalty. In Figure \ref{fig:cv bic}, 5-fold CV selects a quite large $\hat{\lambda} = -0.9$ compared to $\hat{\lambda} = -3$ selected by EBIC (as displayed in Figure 7 in the main text). Actually, $\hat{\lambda} = -3$ selected by EBIC correctly estimates the hierarchy (which is a desirable outcome of the first step), while
$\hat{\lambda} = - 0.9$ selected by CV overestimates the number of permissible attribute patterns. This observation implies that CV fails in the first step to recover the correct attribute hierarchy, which will subsequently lead to an erroneous estimation of LCBN parameters when proceeding to the second step.
In fact, the failure of CV in this setting is in line with some known results regarding CV's model selection inconsistency in regression, e.g.:
``\emph{it is well known from
simulations that the cross-validated Lasso estimator typically selects too many variables}''
\citep[Remark 4.4. in][]{chetverikov2021cross}. 
On the other hand, our selection criterion, the EBIC, has a nice consequence of model selection consistency under high dimensional sparse settings \citep{EBIC}, which justifies using EBIC here. 

\begin{figure}[h!]
    \centering
    \includegraphics[height=4.5cm]{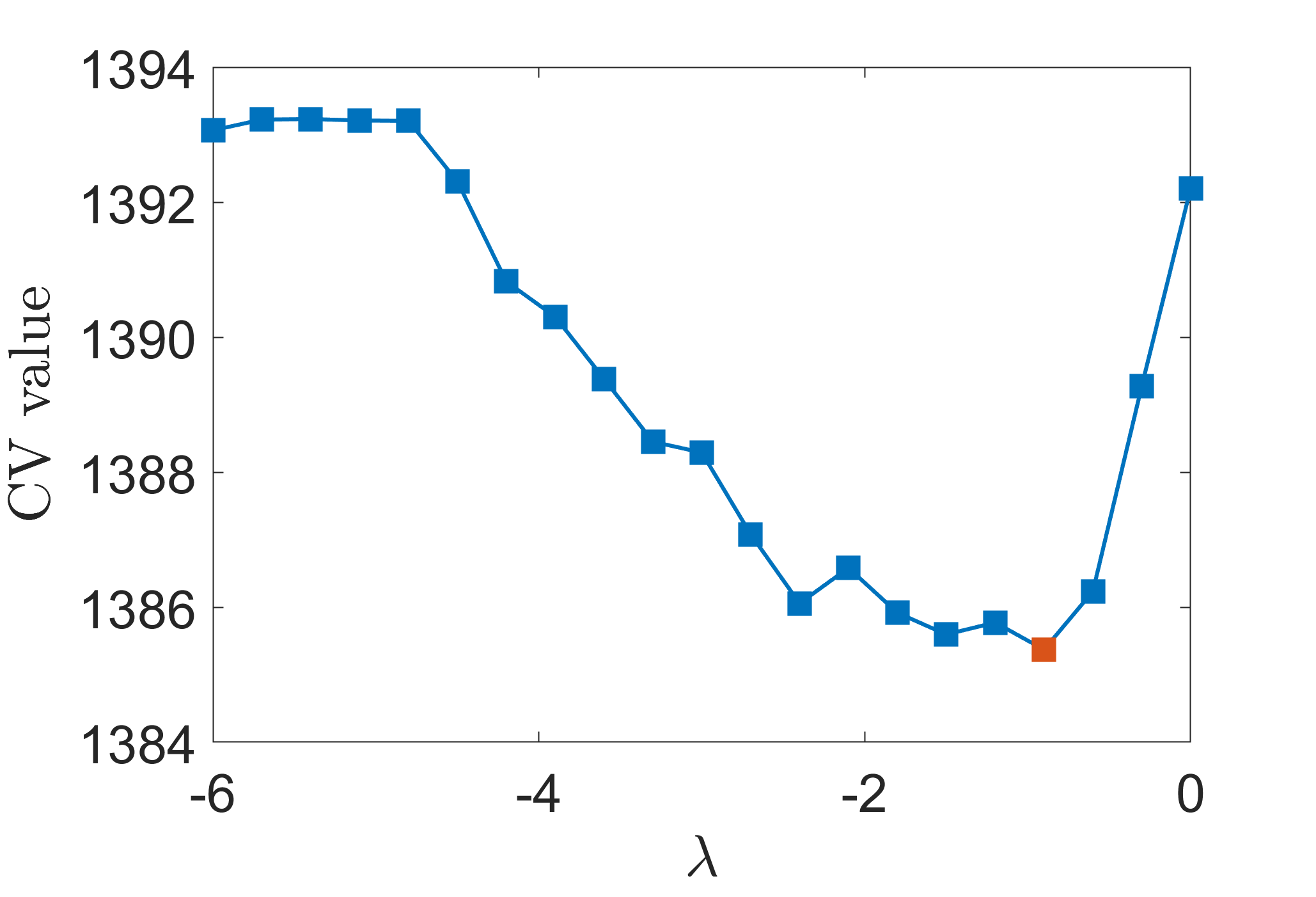}
    \caption{\darkblue{Cross validation value versus $\lambda$. The red point with $\hat{\lambda} = -0.9$ is selected.}} 
    \label{fig:cv bic}
\end{figure}

Additionally, in order to do cross validation, one needs to fit the model multiple times for each fixed value of $\lambda$, which makes it computationally much slower compared to other model selection criteria. Indeed, in our simulations, 5-fold CV took 62.7 seconds whereas using EBIC (or BIC) only took 15.8 seconds.

\color{black}
\subsection{Additional simulations under a misspecified model}
In this section, we present additional simulation results when the model is misspecified, i.e., when the data-generating mechanism does not follow an LCBN. We consider the same diamond hierarchy defined in Figure \ref{fig:diamond}, but instead of constraining the proportion parameters to follow a CBN structure, we initialize $\pp = (\frac{1}{15}, ..., \frac{1}{15})^\top$. In other words, we assign equal probability for the 15 possible patterns in Table \ref{tab:diamond}. %
This initialization follows the attribute hierarchy method, but does not follow an LCBN. Note that this was not the case in the initialization in Section 5.2, which did not follow an attribute hierarchy nor an LCBN.
We compute the pattern selection accuracy Acc$(\hat \mca)$ and the RMSE of parameter estimates in addition to comparing the EBIC and BIC of the first-step estimate and our two-step estimate. The results are summarized in Table \ref{tab-dina-mis-2}.

Table \ref{tab-dina-mis-2} shows that the set of permissible skill patterns $\mca$ is selected successfully most of the time (with a slightly smaller accuracy compared to the results in Section 5.2). Also, the continuous parameters are estimated with an estimation error similar to Table 3 in the main paper, which corresponds to the estimation error of the correctly specified model. 
We also see that except for one scenario ($N = 500, r = 0.3$), the second stage estimate has a lower EBIC and BIC. These results are analogous to those of Section 5.2., and justifies adopting LCBNs even in misspecified scenarios.

\begin{table}[h!]
\centering
\caption{Estimation accuracy and RMSE for the estimated parameters for the misspecified DINA model.}
\label{tab-dina-mis-2}
\begin{tabular}{cccccccc}
\toprule
Model & $N$ & $r$ & Method & Acc($\hat{\mca}$)&  argmin EBIC & argmin BIC & RMSE($\hat{\TT}$)  \\
\midrule 
\multirow{8}{*}{\centering DINA} & \multirow{4}{*}{$500$} &\multirow{2}{*}{$0.2$} 
    &  PEM & - & 3\%  & 11\%  & $0.063$ \\ 
& & & Proposed & 0.52 & 97\%  & 89\%  & $0.057$ \\ 
\cmidrule(lr){3-8}
& &\multirow{2}{*}{$0.3$} 
    &  PEM & - & 48\% & 53\% & $0.097$  \\ 
& & &  Proposed & 0.83 & 52\% & 47\% & $0.079$  \\ 
\cmidrule(lr){2-8} 
& \multirow{4}{*}{$1000$} &\multirow{2}{*}{$0.2$} 
    &  PEM  & - & 5\%  & 28\%  & $0.035$ \\ 
& & &  Proposed & 0.75 & 95\%  & 72\% & $0.023$ \\ 
\cmidrule(lr){3-8} 
& &\multirow{2}{*}{$0.3$} 
    &  PEM  & - & 15\% & 18\% & $0.073$ \\ 
& & &  Proposed & 0.82  & 85\% & 82\% & $0.054$ \\ 
\bottomrule 
\end{tabular}

\end{table}

\subsection{Additional simulation results under an unknown $\QQ$-matrix}

This section presents additional simulation results where the $\QQ$-matrix and the number of latent skills, $K$, are unknown. We apply the exploratory penalized EM algorithm in \cite{ma2022learning} in our first step to jointly estimate the $\QQ$-matrix and the attribute hierarchy $\mce$ (see Section 4.3 in the main text for more details). Our second step for estimating the continuous parameters is still our new Algorithm \ref{algo-em}. In terms of choosing the tuning parameters, we follow the suggestions and settings in \cite{ma2022learning}.

We consider $K = 4$ attributes with a convergent hierarchy in Figure \ref{fig:K=4 convergent}, which is an example hierarchy presented in the initial attribute hierarchy paper \cite{gierl2007ahm}. This convergent hierarchy was also used in the simulation studies in \cite{ma2022learning} but with true proportion parameters that do not follow an LCBN. This hierarchy results in $|\mca| = 6$ permissible patterns, as shown in Table \ref{tab:proportion}.
Here, we set the $K=4$ LCBN parameters as $\bt = (0.9, 0.65, 0.65, 0.5)$ and consider the following $\QQ$-matrix:
$$\QQ = \begin{pmatrix}
    \QQ_1 \\
    \QQ_1 \\
    \QQ_2 \\
    \QQ_2 \\
    \QQ_3 \\
    \mathbf I_K \\
    \mathbf I_K \\
    \mathbf I_K
\end{pmatrix}$$
with $J = 30$ items, where $\QQ_1, \QQ_2$ are the matrices defined in \eqref{eq:example Q} in the main text, $\mathbf I_K$ is the identity matrix, and 
$$\QQ_3 = \begin{pmatrix}
    1 & 0 & 1 & 1 \\
    1 & 1 & 0 & 1
\end{pmatrix}.$$
For the measurement model, we consider the DINA model. In terms of the noise level $r$, we continue to consider $r = 0.1$ and $0.2$ as in the main text.

In Table \ref{tab-dina-exploratory}, we report the accuracy of the estimated hierarchy and the $\QQ$-matrix along with the final estimate of the continuous parameters. In each simulation setting, we perform 100 independent replications and report the average accuracy. 
Here, the accuracy for the $\QQ$-matrix  (matrix-wise or row-wise) is defined in terms of estimating the $\QQ$-matrix up to an equivalence class, as the $\QQ$-matrix under the DINA model is identifiable only up to the equivalence classes defined by the $\Gamma$ matrix \citep{gu2022hlam}. Note that the matrix-wise accuracy (denoted as Acc($\hat{\QQ}^{\text{mat}}$) in Table \ref{tab-dina-exploratory}) is more stringent than the row-wise (or item level) accuracy (denoted as Acc($\hat{\QQ}^{\text{row}}$) in Table \ref{tab-dina-exploratory}) by definition.
As the value and interpretation of the parameter $\bt$ heavily depends on the specific hierarchy and $(\boldsymbol{s}, \boldsymbol{g})$ depends on the $\QQ$-matrix, we compute the RMSE of the continuous parameters when the hierarchy and the $\QQ$-matrix are correctly estimated.
Due to the variety of the tuning parameters that need to be chosen, the runtime for this simulation was much larger than those based on Algorithm \ref{algo-pem} in our main text (it took more than 5 minutes on average when $N = 500$, which is much larger than that reported in Table \ref{tab:time}). 

Table \ref{tab-dina-exploratory} shows that the attribute hierarchy and the $\QQ$-matrix can be jointly estimated with high accuracy, even when the $\QQ$-matrix is unknown. In particular, when the noise level $r$ is small and $N$ is large, we see that both the hierarchy and the $\QQ$-matrix are perfectly estimated, and also the continuous parameters have a small estimation error. Compared to Table \ref{tab-dina} in the main body of our paper, the estimation accuracy of Algorithm \ref{algo-ma} is comparable to Algorithm \ref{algo-pem} when $r = 0.1$ but is much lower when $r = 0.2$ (here, the settings of the two Tables are different due to different numbers of attributes and different hierarchies, so we are only comparing a rough trend of the estimation accuracy). We believe that a larger noise level makes the problem more challenging when the $\QQ$-matrix is unknown. This observation is also coherent with Table 2 in \cite{ma2022learning}. 
}

\bigskip
\begin{minipage}{\textwidth}
    \centering
  \begin{minipage}[c]{0.3\textwidth}
    \centering
    \resizebox{0.8\textwidth}{!}{
    \begin{tikzpicture}[scale=1.2]
	\node (h1)[hiddens] at (1.5, 2) {$\alpha_1$};
    \node (h2)[hiddens] at (0.5, 1) {$\alpha_2$};
    \node (h3)[hiddens] at (2.5, 1) {$\alpha_3$};
    \node (h4)[hiddens] at (1.5, 0) {$\alpha_4$};
    
    
    \draw[pre] (h1) -- (h2); \draw[pre] (h1) -- (h3);
    \draw[pre] (h2) -- (h4); \draw[pre] (h3) -- (h4);
\end{tikzpicture}
}

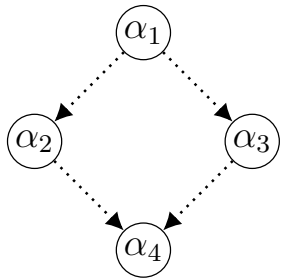
\captionof{figure}{Convergent hierarchy with $K=4$ attributes.}
\label{fig:K=4 convergent}
\end{minipage}
\quad
\begin{minipage}[c]{0.5\textwidth}
\centering
\small
\resizebox{0.9\textwidth}{!}{
\begin{tabular}{c|cccc|c}
\hline
$\mca(\mce)$  &  $\alpha_1$ & $\alpha_2$ & $\alpha_3$ & $\alpha_4$ & $p_{\aaa}$\\
\hline
$\aaa_1$ & 0  &   0  &   0    & 0  & 0.100\\
$\aaa_2$ & 1  &   0  &   0    & 0  & 0.110\\
$\aaa_3$ & 1  &   1  &   0    & 0  & 0.205\\
$\aaa_4$ & 1  &   0  &   1    & 0  & 0.205\\
$\aaa_5$ & 1  &   1  &   1    & 0  & 0.190\\
$\aaa_6$ & 1  &   1  &   1    & 1  & 0.190\\
\hline
\end{tabular}
}
\captionof{table}{Permissible patterns under the convergent hierarchy}
\label{tab:proportion}
\end{minipage}
\end{minipage}

\begin{table}[h!]
\centering
\caption{Estimation accuracy and RMSE for the estimated parameters for the exploratory DINA model.}
\label{tab-dina-exploratory}
\begin{tabular}{cccccccc}
\toprule
Model & $N$ & $r$ (noise level) & Acc($\hat{\mce}$) & Acc($\hat{\QQ}^{\text{mat}}$) & Acc($\hat{\QQ}^{\text{row}}$) & RMSE($\hat{\TT}$) & RMSE($\hat{\bt}$) \\
\midrule 
\multirow{6}{*}{\centering DINA} & \multirow{2}{*}{$500$} &\multirow{1}{*}{$0.1$} 
    & 0.87 & 0.50  & 0.93 & $0.029$ & $0.042$\\
\cmidrule(lr){3-8}
& &\multirow{1}{*}{$0.2$} 
    & 0.14 & 0.00  & 0.83 & $0.046$ & $0.053$\\
\cmidrule(lr){2-8}
& \multirow{2}{*}{$1000$} &\multirow{1}{*}{$0.1$} 
    & 1.00 & 0.94  & 0.99 & $0.021$ & $0.027$\\
    \cmidrule(lr){3-8} 
& &\multirow{1}{*}{$0.2$} 
        & 0.65 & 0.26  & 0.93 & $0.033$ & $0.038$\\
\cmidrule(lr){2-8}
& \multirow{2}{*}{$2000$} &\multirow{1}{*}{$0.1$} 
        & 1.00 & 1.00 & 1.00 & $0.015$ & $0.021$\\
\cmidrule(lr){3-8} 
& &\multirow{1}{*}{$0.2$} 
        & 0.86 & 0.50  & 0.94 & $0.021$ & $0.022$\\
\bottomrule 
\end{tabular}
\end{table}

\subsection{Sensitivity analysis for choosing the tuning parameter $\lambda$ in the log penalty}

We next present simulation evidence to show that the estimation of the hierarchy graph is not very sensitive to the value of $\lambda$.
Consider the same setting as those for the DINA model simulation in the previous subsection. In a simulation trial with sample size $N = 500$ and noise level $r = 0.2$, we plot the number of selected skill patterns and the corresponding EBIC value versus a sequence of $\lambda \in \{0, -0.3, \ldots, -6\}$ in Figure \ref{fig:support size}.  The left panel in Figure \ref{fig:support size} shows a wide interval $\lambda \in [-1.2, -3.6]$ colored in red, with every $\lambda$ in this interval leading to a correct estimate of the attribute hierarchy. 
This fact demonstrates that the estimation of the hierarchy graph $\mce$ is robust to the choice of $\lambda$. Furthermore, even for a stronger penalty with $\lambda \le -3.9$ that is outside of this interval, the estimated hierarchy graph makes only one error by additionally including one prerequisite $\alpha_8 \rightarrow \alpha_7$. 
The right panel in Figure \ref{fig:support size} shows that $\hat{\lambda} = -3$ is chosen via EBIC because it gives smallest EBIC value. 
We can see that EBIC succeeds here because -3 belongs to the feasible interval $[-1.2, -3.6]$.

    \begin{figure}[h!]
        \centering
        \includegraphics[height=4.5cm]{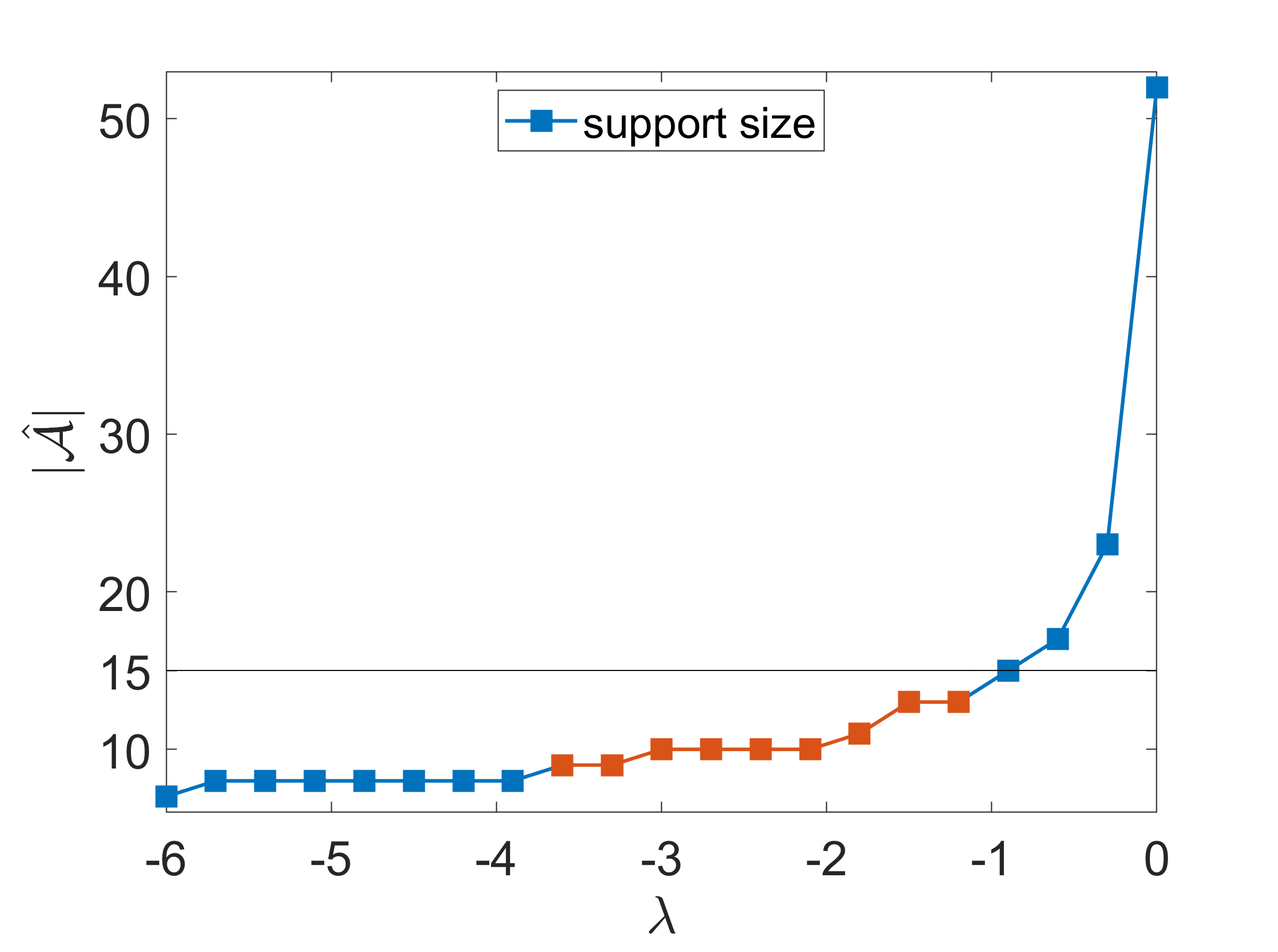}
        \quad
        \includegraphics[height=4.5cm]{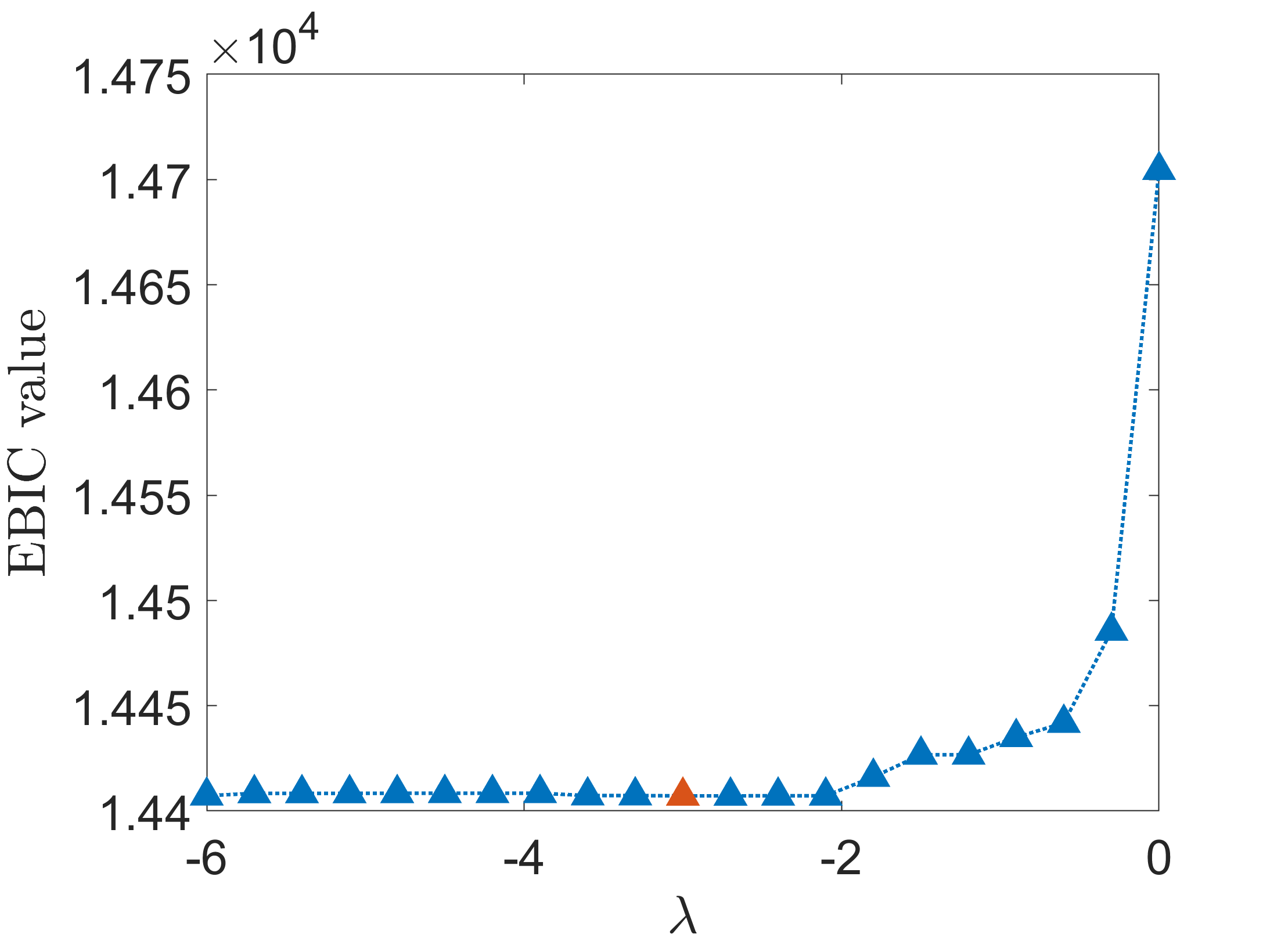}
        \caption{Number of selected skill patterns $|\hat{\mca}|$ (left) and EBIC value (right) plotted against $\lambda$. 
        \textbf{Left}: the red points correspond to a wide interval $[-1.2, -3.6]$ of $\lambda$ that can correctly estimate the hierarchy. \textbf{Right}: $\hat{\lambda} = -3$ is selected via EBIC because it gives the smallest EBIC value.
        }
        \label{fig:support size}
    \end{figure}

\color{black}

\end{document}